\documentclass[epj]{svjour}
\usepackage{graphics}

\begin{document}
\title{Chiral Baryon Fields in the QCD Sum Rule}

\author{Hua-Xing Chen
}                     
\institute{School of Physics and Nuclear Energy Engineering, Beihang University, Beijing 100191, China}
\date{Received: date / Revised version: date}
%
\abstract{We study the structure of local baryon fields using the method of QCD sum rule. We only consider the single baryon fields and calculate their operator product expansions. We find that the octet baryon fields belonging to the chiral representations $[(\mathbf 3,\mathbf{\bar 3}) \oplus (\mathbf{\bar 3}, \mathbf 3)]$ and $[(\mathbf 8,\mathbf{1}) \oplus (\mathbf{1}, \mathbf 8)]$ and the decuplet baryon fields belonging to the chiral representations $[(\mathbf 3,\mathbf 6) \oplus (\mathbf 6, \mathbf 3)]$ lead to the baryon masses which are consistent with the experimental data of ground baryon masses. We also calculate their decay constants, check our normalizations for baryon fields in Ref.~\cite{Chen:2009sf} and find that they are well-defined.
} 
\maketitle

%
\section{Introduction}
\label{intro}

Quantum Chromodynamics (QCD) has an interesting property called the spontaneously breaking of the chiral symmetry. This effect is important to explain the origin of the masses of hadrons. The chiral symmetry itself is important, and has been studied by lots of physicists for more than fifty years, even before QCD became the theory of strong interactions~\cite{Weinberg:1969hw,Hatsuda:1994pi,lee72,lee81,Jido:2001nt,Jido:1997yk,Jido:1999hd,Beane:2002td,Benmerrouche:1989uc,Haberzettl:1998rw,DeTar:1988kn,Cohen:2002st,Diakonov:1987ty,Nagata:2007di}.

In Ref.~\cite{Weinberg:1969hw}, S.~Weinberg pointed out that the algebraic aspects of chiral symmetry are worth studying. Following his idea, we have classified the baryon fields which are composed by three quarks~\cite{Chen:2008qv}. We note here that some identities for baryon currents have been studied by B.~L.~Ioffe, Y.~Chung, D.~Espriu and D.~B.~Leinweber, etc.~\cite{Ioffe:1981kw,Chung:1981cc,Espriu:1983hu,Leinweber:1995ie,Leinweber:1994nm}. To use the quark fields, we have considered their flavor structure, color structure and the Lorentz structure, where the chiral symmetry can be naturally included. We have verified that there are altogether one flavor singlet baryon field, three flavor octet baryon fields and two flavor decuplet baryon fields in the limit of local interpolating fields. These fields are independent as well as complete to describe all the local three-quark baryon fields. The possible chiral representations are $[(\mathbf 3,\mathbf{\bar 3}) \oplus (\mathbf{\bar 3}, \mathbf 3)]$ (containing one flavor singlet field and one flavor octet field), $[(\mathbf 8,\mathbf 1) \oplus (\mathbf 1, \mathbf 8)]$ (containing one flavor octet field), $[(\mathbf 6,\mathbf 3) \oplus (\mathbf 3, \mathbf 6)]$ (containing one flavor octet field and one flavor decuplet field) and $[(\mathbf{10},\mathbf 1) \oplus (\mathbf 1, \mathbf{10})]$ (containing one flavor decuplet field). We have also studied the axial coupling constant $g_A$ and the masses due to the mixing of different chiral multiplets using a ``simple'' algebraic method~\cite{Chen:2008qv}.

In this paper we try to study the masses of the baryon ground states, using these baryon fields but with a different method, the method of the QCD sum rule. Since all the local baryon fields have been classified in Ref.~\cite{Chen:2008qv}, we can use them straightforwardly to perform the QCD sum rule analysis. The baryon state can be decomposed in terms of a Fock space expansion as three quarks ($|qqq\rangle$), five quarks ($|qqqq \bar q\rangle$), etc. Even for the three-quark components, there can be more than one structures. For example, three are three independent octet baryon fields, all of which can couple to the octet baryon states, like proton. A proper combination of these fields may couple to the baryon ground octet state in an optimize way. We have also found in Ref.~\cite{Chen:2009sf} that the mixing of different chiral representations is important to explain the physical baryon masses. However, the single baryon field itself can also couple to the relevant baryon state, as long as it exists in the Fock space expansion. Therefore, as a first step, in this paper we use the single baryon fields to perform the QCD sum rule analysis. It is our next task to study their mixing . We note that this subject has been studied by many authors, but the baryon fields used are different~\cite{Ioffe:1981kw,Chung:1981cc,Espriu:1983hu,Lee:2002jb,Leinweber:1995fn,He:1996wy,Dai:1996xv,Yang:1993bp,Furnstahl:1992pi,Dosch:1988vv}. Those used in the Lattice QCD calculations are also different~\cite{Leinweber:2004it,Zanotti:2003fx,Fucito:1982ip}.

This paper is organized as follows. In Sec.~\ref{sec:field} we rewrite the baryon fields of Ref.~\cite{Chen:2008qv} to a proper form which we can use to perform the QCD sum rule analysis in the next section, Sec.~\ref{sec:svz}. The full results of operator product expansion (OPE) are shown in Appendix.~\ref{app:ope}, and the simplified results are shown in Sec.~\ref{sec:ope}, which are in the $SU(3)$ limit and after assuming $m_q = 0$. In Sec.~\ref{sec:numerical} the numerical results are obtained for their masses. Sec.~\ref{sec:summary} is a summary.

\section{Baryon Fields}
\label{sec:field}

In Ref.~\cite{Chen:2008qv,Chen:2009sf} we have classified the baryon interpolating fields as local products of three quarks according to their flavor structure and chiral structure. Here we use the chiral symmetry group $SU(3)_L \otimes SU(3)_R$ to show our previous results and transform them to a proper form which we can use to perform the QCD sum rule analysis. We note that the notations and normalizations both follow Ref.~\cite{Chen:2009sf}.

The flavor representations of baryon fields composed by three quarks can be singlet, octet and decuplet:
\begin{eqnarray}
\mathbf 3 \otimes \mathbf 3 \otimes \mathbf 3 = \mathbf 1 \oplus \mathbf 8 \oplus \mathbf 8 \oplus \mathbf{10} \, ,
\end{eqnarray}
while the revelent chiral representations can be $[(\mathbf 1,\mathbf 1) \oplus (\mathbf 1, \mathbf 1)]$, $[(\mathbf 3,\mathbf{\bar 3}) \oplus (\mathbf{\bar 3}, \mathbf 3)]$, $[(\mathbf 8,\mathbf 1) \oplus (\mathbf 1, \mathbf 8)]$, $[(\mathbf 6,\mathbf 3) \oplus (\mathbf 3, \mathbf 6)]$ and $[(\mathbf{10},\mathbf 1) \oplus (\mathbf 1, \mathbf{10})]$:
\begin{eqnarray}
&& [(\mathbf 3,\mathbf 1) \oplus (\mathbf 1, \mathbf 3)]^3
\\ \nonumber &\ni& [(\mathbf 1,\mathbf 1) \oplus (\mathbf 1, \mathbf 1)] \oplus [(\mathbf 3,\mathbf{\bar 3}) \oplus (\mathbf{\bar 3}, \mathbf 3)] \oplus [(\mathbf 8,\mathbf 1) \oplus (\mathbf 1, \mathbf 8)]
\\ \nonumber && \oplus [(\mathbf 6,\mathbf 3) \oplus (\mathbf 3, \mathbf 6)] \oplus [(\mathbf{10},\mathbf 1) \oplus (\mathbf 1, \mathbf{10})] \, .
\end{eqnarray}
In Ref.~\cite{Chen:2008qv} we have found that there is one baryon field for each chiral multiplet except $[(\mathbf 1,\mathbf 1) \oplus (\mathbf 1, \mathbf 1)]$. They are Dirac spinor fields:
\begin{eqnarray}
\nonumber J^{(\mathbf 3,\mathbf{\bar 3}) \rightarrow \mathbf{1}_f} &\equiv& \Lambda = {2\sqrt2\over\sqrt3} \epsilon_{abc} \epsilon^{ABC} (q^{aT}_A C q^b_B) \gamma_5 q^c_C \, ,
\\ \nonumber J^{(\mathbf 3,\mathbf{\bar 3}) \rightarrow \mathbf{8}_f} &\equiv& N_-^N = N_1^N - N_2^N \, ,
\\ J^{(\mathbf 8,\mathbf 1) \rightarrow \mathbf{8}_f} &\equiv& N_+^N = N_1^N + N_2^N \, ,
\end{eqnarray}
the Rarita-Schwinger's vector-spinor field:
\begin{eqnarray}
\nonumber J^{(\mathbf 6,\mathbf 3) \rightarrow \mathbf{8}_f} &\equiv& N_\mu^N = \epsilon_{abc} \epsilon^{ABD} \lambda_{DC}^N (q^{aT}_A C \gamma_\mu \gamma_5 q^b_B) \gamma_5 q^c_C
\\ && ~~~~~~~ + {1\over4} \gamma_\mu \gamma_5 (N^N_1 - N^N_2) \, ,
\\ \nonumber J^{(\mathbf 6,\mathbf 3) \rightarrow \mathbf{10}_f} &\equiv& \Delta^P_{\mu} = \epsilon_{abc} S_P^{ABC} (q^{aT}_A C \gamma_\mu q^b_B) q^c_C \, ,
\end{eqnarray}
and the antisymmetric-tensor-spinor field:
\begin{eqnarray}
J^{(\mathbf{10},\mathbf 1) \rightarrow \mathbf{10}_f} &\equiv& \Delta^P_{\mu\nu} = \epsilon_{abc} S_P^{ABC} (q^{aT}_A C \sigma_{\mu\nu} q^b_B) \gamma_5 q^c_C
\\ \nonumber && ~~~~~~~~ - {i\over2} \gamma_\mu \gamma_5 \Delta^P_{\nu} + {i\over2} \gamma_\nu \gamma_5 \Delta^P_{\mu} \, ,
\end{eqnarray}
where $N^N_1$ and $N^N_2$ are
\begin{eqnarray}
N^N_1 &=&
\epsilon_{abc} \epsilon^{ABD} \lambda_{DC}^N (q^{aT}_A C q^b_B) \gamma_5 q^c_C \, ,
\\ \nonumber N^N_2 &=& \epsilon_{abc} \epsilon^{ABD} \lambda_{DC}^N (q^{aT}_A C \gamma_5 q^b_B) q^c_C \, .
\end{eqnarray}
These interpolating fields are composed by three quark fields. The quark field $q$ is a {\it flavor} triplet and contains a {\it color} index:
\begin{equation}
q^a_A \in {\bf q}^a = (u^a \, , \, d^a \, , \, s^a)^{\rm T} \, ;
\end{equation}
In these baryon fields the summation is taken over repeated indices ($a$, $b$, $c$ for color indices and $A$, $B$, $\cdots$ for flavor indices). Indices $a$, $A$, $\cdots$ take values $1\cdots3$, $N$ takes values $1\cdots8$, and $P$ takes values $1\cdots10$. $S_P^{ABC}$ is the totally symmetric tensor, $\epsilon^{ABC}$ is the totally anti-symmetric tensor and $\bf \lambda^N$ is the Gell-Mann matrices.

To perform the QCD sum rule analysis, we need to sum over flavor indices. We show the explicit forms of these interpolating fields in the following subsections according to their chiral representations. We find that a simple baryon field always contains two or more chiral components, for example, $\epsilon_{abc} (u^T_a C d_b) \gamma_5 u_c$ contains both $[(\mathbf 3,\mathbf{\bar 3}) \oplus (\mathbf{\bar 3}, \mathbf 3)]$ and $[(\mathbf 8,\mathbf 1) \oplus (\mathbf 1, \mathbf 8)]$ components. The three exceptions are $\epsilon^{abc} (u^{T}_a C \gamma_\mu u_b) u_c$, $\epsilon^{abc} (d^{T}_a C \gamma_\mu d_b) d_c$ and $\epsilon^{abc} (s^{T}_a C \gamma_\mu s_b) s_c$ which all belong to the chiral representation $[(\mathbf 6,\mathbf 3) \oplus (\mathbf 3, \mathbf 6)]$.

\subsection{Chiral Multiplet $[(\mathbf 3,\mathbf{\bar 3}) \oplus (\mathbf{\bar 3}, \mathbf 3)]$}

We use $\eta$ to denote the baryon fields belonging to the chiral multiplet $[(\mathbf 3,\mathbf{\bar 3}) \oplus (\mathbf{\bar 3}, \mathbf 3)]$, which contains one flavor singlet and one flavor octet. The interpolating field for the singlet baryon $\Lambda_1$ is
\begin{eqnarray}
&& \eta^{\Lambda_1} \equiv \Lambda
\\ \nonumber &=& {2\sqrt2\over\sqrt3} \epsilon_{abc} \epsilon^{ABC} (q^{aT}_A C q^b_B) \gamma_5 q^c_C
\\ \nonumber &=& {4\sqrt2\over\sqrt3} \epsilon^{abc} \big ( u^{T}_a C d_b \gamma_5 s_c - u^{T}_a C s_b \gamma_5 d_c + d^{T}_a C s_b \gamma_5 u_c \big ) \, .
\end{eqnarray}
The interpolating fields for the octet baryons belonging to the chiral representation $[(\mathbf 3,\mathbf{\bar 3}) \oplus (\mathbf{\bar 3}, \mathbf 3)]$ are
\begin{eqnarray}
\nonumber \eta^{p} &\equiv& N_-^p = {N^4_1 - i N^5_1 \over \sqrt 2} - { N^4_2 - i N^5_2 \over \sqrt 2}
\\ \nonumber && ~~~~ = 2 \sqrt 2 \epsilon_{abc} \big ( u^T_a C d_b \gamma_5 u_c - u^T_a C \gamma_5 d_b u_c \big ) \, ,
\\ \nonumber \eta^{n} &\equiv& N_-^{n} = 2 \sqrt 2 \epsilon_{abc} \big ( u^T_a C d_b \gamma_5 d_c - u^T_a C \gamma_5 d_b d_c \big ) \, ,
\\ \nonumber \eta^{\Sigma^+} &\equiv& N_-^{\Sigma^+} = 2 \sqrt 2 \epsilon_{abc} \big ( u^T_a C s_b \gamma_5 u_c - u^T_a C \gamma_5 s_b u_c \big ) \, ,
\\ \eta^{\Sigma^0} &\equiv& N_-^{\Sigma^0} = 2 \epsilon_{abc} \big ( d^T_a C s_b \gamma_5 u_c + u^T_a C s_b \gamma_5 d_c
\\ \nonumber && ~~~~~~~~ - d^T_a C \gamma_5 s_b u_c - u^T_a C \gamma_5 s_b d_c \big ) \, ,
\\ \nonumber \eta^{\Sigma^-} &\equiv& N_-^{\Sigma^-} = 2 \sqrt 2 \epsilon_{abc} \big ( d^T_a C s_b \gamma_5 d_c - d^T_a C \gamma_5 s_b d_c \big ) \, ,
\\ \nonumber \eta^{\Xi^0} &\equiv& N_-^{\Xi^0} = - 2 \sqrt 2 \epsilon_{abc} \big ( u^T_a C s_b \gamma_5 s_c - u^T_a C \gamma_5 s_b s_c \big ) \, ,
\\ \nonumber \eta^{\Xi^-} &\equiv& N_-^{\Xi^-} = - 2 \sqrt 2 \epsilon_{abc} \big ( d^T_a C s_b \gamma_5 s_c - d^T_a C \gamma_5 s_b s_c \big ) \, ,
\\ \nonumber \eta^{\Lambda_8} &\equiv& N_-^{\Lambda_8} = {1 \over \sqrt{3}} \epsilon_{abc} \big ( 2 d^T_a C s_b \gamma_5 u_c - 2 u^T_a C s_b \gamma_5 d_c
\\ \nonumber && ~~~~~~~~ - 4 u^T_a C d_b \gamma_5 s_c - 2 d^T_a C \gamma_5 s_b u_c
\\ \nonumber && ~~~~~~~~ + 2 u^T_a C \gamma_5 s_b d_c + 4 u^T_a C \gamma_5 d_b s_c \big ) \, .
\end{eqnarray}

\subsection{Chiral Multiplet $[(\mathbf 8,\mathbf 1) \oplus (\mathbf 1, \mathbf 8)]$}

We use $\xi$ to denote the octet baryon fields belonging to the chiral representation $[(\mathbf 8,\mathbf 1) \oplus (\mathbf 1, \mathbf 8)]$. The interpolating fields for the octet baryons belonging to this chiral representation are
\begin{eqnarray}
\nonumber \xi^{p} &\equiv& N_+^{p} = {N^4_1 - i N^5_1 \over \sqrt 2} + {N^4_2 - i N^5_2 \over \sqrt 2}
\\ \nonumber && ~~~~ = 2 \sqrt 2 \epsilon_{abc} \big ( u^T_a C d_b \gamma_5 u_c + u^T_a C \gamma_5 d_b u_c \big ) \, ,
\\ \nonumber \xi^{n} &\equiv& N_+^{n} = 2 \sqrt 2 \epsilon_{abc} \big ( u^T_a C d_b \gamma_5 d_c + u^T_a C \gamma_5 d_b d_c \big ) \, ,
\\ \nonumber \xi^{\Sigma^+} &\equiv& N_+^{\Sigma^+} = 2 \sqrt 2 \epsilon_{abc} \big ( u^T_a C s_b \gamma_5 u_c + u^T_a C \gamma_5 s_b u_c \big ) \, ,
\\ \xi^{\Sigma^0} &\equiv& N_+^{\Sigma^0} = 2 \epsilon_{abc} \big ( d^T_a C s_b \gamma_5 u_c + u^T_a C s_b \gamma_5 d_c
\\ \nonumber && ~~~~~~~~ + d^T_a C \gamma_5 s_b u_c + u^T_a C \gamma_5 s_b d_c \big ) \, ,
\\ \nonumber \xi^{\Sigma^-} &\equiv& N_+^{\Sigma^-} = 2 \sqrt 2 \epsilon_{abc} \big (  d^T_a C s_b \gamma_5 d_c + d^T_a C \gamma_5 s_b d_c \big ) \, ,
\\ \nonumber \xi^{\Xi^0} &\equiv& N_+^{\Xi^0} = - 2 \sqrt 2 \epsilon_{abc} \big ( u^T_a C s_b \gamma_5 s_c + u^T_a C \gamma_5 s_b s_c \big ) \, ,
\\ \nonumber \xi^{\Xi^-} &\equiv& N_+^{\Xi^-} = - 2 \sqrt 2 \epsilon_{abc} \big ( d^T_a C s_b \gamma_5 s_c + d^T_a C \gamma_5 s_b s_c \big ) \, ,
\\ \nonumber \xi^{\Lambda_8} &\equiv& N_+^{\Lambda_8} = {1 \over \sqrt{3}} \epsilon_{abc} \big ( 2 d^T_a C s_b \gamma_5 u_c - 2 u^T_a C s_b \gamma_5 d_c
\\ \nonumber && ~~~~~~~~ - 4 u^T_a C d_b \gamma_5 s_c + 2 d^T_a C \gamma_5 s_b u_c
\\ \nonumber && ~~~~~~~~ - 2 u^T_a C \gamma_5 s_b d_c - 4 u^T_a C \gamma_5 d_b s_c \big ) \, .
\end{eqnarray}

\subsection{Chiral Multiplet $[(\mathbf 6,\mathbf 3) \oplus (\mathbf 3, \mathbf 6)]$}

We use $\psi$ to denote the baryon fields belonging to the chiral representation $[(\mathbf 6,\mathbf 3) \oplus (\mathbf 3, \mathbf 6)]$, which contains one flavor octet and one flavor decuplet. The interpolating fields for the octet baryons belonging to the chiral representation $[(\mathbf 6,\mathbf 3) \oplus (\mathbf 3, \mathbf 6)]$ are
\begin{eqnarray}
\nonumber \psi_\mu^{p} &\equiv& N_\mu^{p} = {N^4_\mu - i N^5_\mu \over \sqrt 2}
\\ \nonumber && ~~~~ = {\epsilon_{abc} \over \sqrt2}  \big ( 4 u^T_a C \gamma_\mu \gamma_5 d_b \gamma_5 u_c + u^T_a C d_b \gamma_\mu u_c
\\ \nonumber && ~~~~~~~~ - u^T_a C \gamma_5 d_b \gamma_\mu \gamma_5 u_c \big ) \, ,
\\ \nonumber \psi_\mu^{n} &\equiv& N_\mu^{n} = {\epsilon_{abc} \over \sqrt2}  \big ( 4 u^T_a C \gamma_\mu \gamma_5 d_b \gamma_5 d_c + u^T_a C d_b \gamma_\mu d_c
\\ \nonumber && ~~~~~~~~ - u^T_a C \gamma_5 d_b \gamma_\mu \gamma_5 d_c \big ) \, ,
\\ \nonumber \psi_\mu^{\Sigma^+} &\equiv& N_\mu^{\Sigma^+} = {\epsilon_{abc} \over \sqrt2}  \big ( 4 u^T_a C \gamma_\mu \gamma_5 s_b \gamma_5 u_c + u^T_a C s_b \gamma_\mu u_c
\\ \nonumber && ~~~~~~~~ - u^T_a C \gamma_5 s_b \gamma_\mu \gamma_5 u_c \big ) \, ,
\\ \psi_\mu^{\Sigma^0} &\equiv& N_\mu^{\Sigma^0} = {\epsilon_{abc} \over 2}  \big ( 4 d^T_a C \gamma_\mu \gamma_5 s_b \gamma_5 u_c + d^T_a C s_b \gamma_\mu u_c
\\ \nonumber && ~~~~~~~~ - d^T_a C \gamma_5 s_b \gamma_\mu \gamma_5 u_c + 4 u^T_a C \gamma_\mu \gamma_5 s_b \gamma_5 d_c
\\ \nonumber && ~~~~~~~~ + u^T_a C s_b \gamma_\mu d_c - u^T_a C \gamma_5 s_b \gamma_\mu \gamma_5 d_c \big ) \, ,
\\ \nonumber \psi_\mu^{\Sigma^-} &\equiv& N_\mu^{\Sigma^-} = {\epsilon_{abc} \over \sqrt2}  \big ( 4 d^T_a C \gamma_\mu \gamma_5 s_b \gamma_5 d_c + d^T_a C s_b \gamma_\mu d_c
\\ \nonumber && ~~~~~~~~ - d^T_a C \gamma_5 s_b \gamma_\mu \gamma_5 d_c \big ) \, ,
\\ \nonumber \psi_\mu^{\Xi^0} &\equiv& N_\mu^{\Xi^0} = {\epsilon_{abc} \over \sqrt2}  \big ( - 4 u^T_a C \gamma_\mu \gamma_5 s_b \gamma_5 s_c - u^T_a C s_b \gamma_\mu s_c
\\ \nonumber && ~~~~~~~~ + u^T_a C \gamma_5 s_b \gamma_\mu \gamma_5 s_c \big ) \, ,
\\ \nonumber \psi_\mu^{\Xi^-} &\equiv& N_\mu^{\Xi^-} = {\epsilon_{abc} \over \sqrt2}  \big ( - 4 d^T_a C \gamma_\mu \gamma_5 s_b \gamma_5 s_c - d^T_a C s_b \gamma_\mu s_c
\\ \nonumber && ~~~~~~~~ + d^T_a C \gamma_5 s_b \gamma_\mu \gamma_5 s_c \big ) \, ,
\\ \nonumber \psi_\mu^{\Lambda_8} &\equiv& N_\mu^{\Lambda_8} = {\epsilon_{abc} \over 2\sqrt3}  \big ( 4 d^T_a C \gamma_\mu \gamma_5 s_b \gamma_5 u_c + d^T_a C s_b \gamma_\mu u_c
\\ \nonumber &-& d^T_a C \gamma_5 s_b \gamma_\mu \gamma_5 u_c - 4 u^T_a C \gamma_\mu \gamma_5 s_b \gamma_5 d_c
\\ \nonumber &-& u^T_a C s_b \gamma_\mu d_c + u^T_a C \gamma_5 s_b \gamma_\mu \gamma_5 d_c - 8 u^T_a C \gamma_\mu \gamma_5 d_b \gamma_5 s_c
\\ \nonumber &-& 2 u^T_a C d_b \gamma_\mu s_c + 2 u^T_a C \gamma_5 d_b \gamma_\mu \gamma_5 s_c \big ) \, .
\end{eqnarray}
The interpolating fields for the decuplet baryons belonging to the chiral representation $[(\mathbf 6,\mathbf 3) \oplus (\mathbf 3, \mathbf 6)]$ are
\begin{eqnarray}
\nonumber \psi_\mu^{\Delta^{++}} &\equiv& \Delta_\mu^1 = -{\sqrt3} \epsilon^{abc} u^{T}_a C \gamma_\mu u_b u_c \, ,
\\ \nonumber \psi_\mu^{\Delta^{+}} &\equiv& \Delta_\mu^2 = - \epsilon^{abc} u^{T}_a C \gamma_\mu u_b d_c - 2 \epsilon^{abc} u^{T}_a C \gamma_\mu d_b u_c \, ,
\\ \nonumber \psi_\mu^{\Delta^{0}} &\equiv& \Delta_\mu^4 = - 2 \epsilon^{abc} u^{T}_a C \gamma_\mu d_b d_c - \epsilon^{abc} d^{T}_a C \gamma_\mu d_b u_c \, ,
\\ \nonumber \psi_\mu^{\Delta^{-}} &\equiv& \Delta_\mu^7 = -{\sqrt3} \epsilon^{abc} d^{T}_a C \gamma_\mu d_b d_c \, ,
\\ \psi_\mu^{\Sigma^{*+}} &\equiv& \Delta_\mu^3 = - \epsilon^{abc} u^{T}_a C \gamma_\mu u_b s_c - 2 \epsilon^{abc} u^{T}_a C \gamma_\mu s_b u_c \, ,
\\ \nonumber \psi_\mu^{\Sigma^{*0}} &\equiv& \Delta_\mu^5 = - \sqrt 2 \epsilon^{abc} u^{T}_a C \gamma_\mu d_b s_c - \sqrt 2 \epsilon^{abc} u^{T}_a C \gamma_\mu s_b d_c
\\ \nonumber && ~~~~~~~~ - \sqrt 2 \epsilon^{abc} d^{T}_a C \gamma_\mu s_b u_c \, ,
\\ \nonumber \psi_\mu^{\Sigma^{*-}} &\equiv& \Delta_\mu^8 = - \epsilon^{abc} d^{T}_a C \gamma_\mu d_b s_c - 2 \epsilon^{abc} d^{T}_a C \gamma_\mu s_b d_c \, ,
\\ \nonumber \psi_\mu^{\Xi^{0}} &\equiv& \Delta_\mu^6 = - 2 \epsilon^{abc} u^{T}_a C \gamma_\mu s_b s_c - \epsilon^{abc} s^{T}_a C \gamma_\mu s_b u_c \, ,
\\ \nonumber \psi_\mu^{\Xi^{-}} &\equiv& \Delta_\mu^9 = - 2 \epsilon^{abc} d^{T}_a C \gamma_\mu s_b s_c - \epsilon^{abc} s^{T}_a C \gamma_\mu s_b d_c \, ,
\\ \nonumber \psi_\mu^{\Omega} &\equiv& \Delta_\mu^{10} = -{\sqrt3} \epsilon^{abc} s^{T}_a C \gamma_\mu s_b s_c \, ,
\end{eqnarray}

\subsection{Chiral Multiplet $[(\mathbf{10},\mathbf 1) \oplus (\mathbf 1, \mathbf{10})]$}

We use $\phi$ to denote the decuplet baryon fields belonging to the chiral representation $[(\mathbf{10},\mathbf 1) \oplus (\mathbf 1, \mathbf{10})]$. The interpolating fields for the decuplet baryons belonging to this chiral representation are
\begin{eqnarray}\label{eq:ope}
\nonumber && -{1\over\sqrt3} \phi_{\mu\nu}^{\Delta^{++}} \equiv -{1\over\sqrt3} \Delta_{\mu\nu}^1 = \epsilon^{abc} u^{T}_a C \sigma_{\mu\nu} u_b \gamma_5 u_c
\\ \nonumber && ~~~~~~~~ - {i\over2} \epsilon^{abc} u^{T}_a C \gamma_\nu u_b \gamma_\mu \gamma_5 u_c + {i\over2} \epsilon^{abc} u^{T}_a C \gamma_\mu u_b \gamma_\nu \gamma_5 u_c \, ,
\\ \nonumber && - \phi_{\mu\nu}^{\Delta^{+}} \equiv - \Delta_{\mu\nu}^2 = \epsilon^{abc} u^{T}_a C \sigma_{\mu\nu} u_b \gamma_5 d_c
\\ \nonumber && ~~~~~~~~ - {i\over2} \epsilon^{abc} u^{T}_a C \gamma_\nu u_b \gamma_\mu \gamma_5 d_c + {i\over2} \epsilon^{abc} u^{T}_a C \gamma_\mu u_b \gamma_\nu \gamma_5 d_c
\\ \nonumber && ~~~~~~~~ + 2 \epsilon^{abc} u^{T}_a C \sigma_{\mu\nu} d_b \gamma_5 u_c - {i} \epsilon^{abc} u^{T}_a C \gamma_\nu d_b \gamma_\mu \gamma_5 u_c
\\ \nonumber && ~~~~~~~~ + {i} \epsilon^{abc} u^{T}_a C \gamma_\mu d_b \gamma_\nu \gamma_5 u_c \, ,
\\ \nonumber && - \phi_{\mu\nu}^{\Delta^{0}} \equiv - \Delta_{\mu\nu}^4 = 2 \epsilon^{abc} u^{T}_a C \sigma_{\mu\nu} d_b \gamma_5 d_c
\\ \nonumber && ~~~~~~~~ - {i} \epsilon^{abc} u^{T}_a C \gamma_\nu d_b \gamma_\mu \gamma_5 d_c + {i} \epsilon^{abc} u^{T}_a C \gamma_\mu d_b \gamma_\nu \gamma_5 d_c
\\ \nonumber && ~~~~~~~~ + \epsilon^{abc} d^{T}_a C \sigma_{\mu\nu} d_b \gamma_5 u_c - {i\over2} \epsilon^{abc} d^{T}_a C \gamma_\nu d_b \gamma_\mu \gamma_5 u_c
\\ \nonumber && ~~~~~~~~ + {i\over2} \epsilon^{abc} d^{T}_a C \gamma_\mu d_b \gamma_\nu \gamma_5 u_c \, ,
\\ \nonumber && -{1\over\sqrt3} \phi_{\mu\nu}^{\Delta^{-}} \equiv -{1\over\sqrt3} \Delta_{\mu\nu}^7 = \epsilon^{abc} d^{T}_a C \sigma_{\mu\nu} d_b \gamma_5 d_c
\\ \nonumber && ~~~~~~~~ - {i\over2} \epsilon^{abc} d^{T}_a C \gamma_\nu d_b \gamma_\mu \gamma_5 d_c + {i\over2} \epsilon^{abc} d^{T}_a C \gamma_\mu d_b \gamma_\nu \gamma_5 d_c \, ,
\\ && - \phi_{\mu\nu}^{\Sigma^{*+}} \equiv - \Delta_{\mu\nu}^3 = \epsilon^{abc} u^{T}_a C \sigma_{\mu\nu} u_b \gamma_5 s_c
\\ \nonumber && ~~~~~~~~ - {i\over2} \epsilon^{abc} u^{T}_a C \gamma_\nu u_b \gamma_\mu \gamma_5 s_c + {i\over2} \epsilon^{abc} u^{T}_a C \gamma_\mu u_b \gamma_\nu \gamma_5 s_c
\\ \nonumber && ~~~~~~~~ + 2 \epsilon^{abc} u^{T}_a C \sigma_{\mu\nu} s_b \gamma_5 u_c - {i} \epsilon^{abc} u^{T}_a C \gamma_\nu s_b \gamma_\mu \gamma_5 u_c
\\ \nonumber && ~~~~~~~~ + {i} \epsilon^{abc} u^{T}_a C \gamma_\mu s_b \gamma_\nu \gamma_5 u_c \, ,
\\ \nonumber && -{\sqrt2} \phi_{\mu\nu}^{\Sigma^{*0}} \equiv -{\sqrt2} \Delta_{\mu\nu}^5 = 2 \epsilon^{abc} u^{T}_a C \sigma_{\mu\nu} d_b \gamma_5 s_c
\\ \nonumber && ~~~~~~~~ - {i} \epsilon^{abc} u^{T}_a C \gamma_\nu d_b \gamma_\mu \gamma_5 s_c + {i} \epsilon^{abc} u^{T}_a C \gamma_\mu d_b \gamma_\nu \gamma_5 s_c
\\ \nonumber && ~~~~~~~~ + 2 \epsilon^{abc} u^{T}_a C \sigma_{\mu\nu} s_b \gamma_5 d_c - {i} \epsilon^{abc} u^{T}_a C \gamma_\nu s_b \gamma_\mu \gamma_5 d_c
\\ \nonumber && ~~~~~~~~ + {i} \epsilon^{abc} u^{T}_a C \gamma_\mu s_b \gamma_\nu \gamma_5 d_c + 2 \epsilon^{abc} d^{T}_a C \sigma_{\mu\nu} s_b \gamma_5 u_c
\\ \nonumber && ~~~~~~~~ - {i} \epsilon^{abc} d^{T}_a C \gamma_\nu s_b \gamma_\mu \gamma_5 u_c + {i} \epsilon^{abc} d^{T}_a C \gamma_\mu s_b \gamma_\nu \gamma_5 u_c \, ,
\\ \nonumber && - \phi_{\mu\nu}^{\Sigma^{*-}} \equiv - \Delta_{\mu\nu}^8 = \epsilon^{abc} d^{T}_a C \sigma_{\mu\nu} d_b \gamma_5 s_c
\\ \nonumber && ~~~~~~~~ - {i\over2} \epsilon^{abc} d^{T}_a C \gamma_\nu d_b \gamma_\mu \gamma_5 s_c + {i\over2} \epsilon^{abc} d^{T}_a C \gamma_\mu d_b \gamma_\nu \gamma_5 s_c
\\ \nonumber && ~~~~~~~~ + 2 \epsilon^{abc} d^{T}_a C \sigma_{\mu\nu} s_b \gamma_5 d_c - {i} \epsilon^{abc} d^{T}_a C \gamma_\nu s_b \gamma_\mu \gamma_5 d_c
\\ \nonumber && ~~~~~~~~ + {i} \epsilon^{abc} d^{T}_a C \gamma_\mu s_b \gamma_\nu \gamma_5 d_c \, ,
\\ \nonumber && - \phi_{\mu\nu}^{\Xi^{0}} \equiv - \Delta_{\mu\nu}^6 = 2 \epsilon^{abc} u^{T}_a C \sigma_{\mu\nu} s_b \gamma_5 s_c
\\ \nonumber && ~~~~~~~~ - {i} \epsilon^{abc} u^{T}_a C \gamma_\nu s_b \gamma_\mu \gamma_5 s_c + {i} \epsilon^{abc} u^{T}_a C \gamma_\mu s_b \gamma_\nu \gamma_5 s_c
\\ \nonumber && ~~~~~~~~ + \epsilon^{abc} s^{T}_a C \sigma_{\mu\nu} s_b \gamma_5 u_c - {i\over2} \epsilon^{abc} s^{T}_a C \gamma_\nu s_b \gamma_\mu \gamma_5 u_c
\\ \nonumber && ~~~~~~~~ + {i\over2} \epsilon^{abc} s^{T}_a C \gamma_\mu s_b \gamma_\nu \gamma_5 u_c \, ,
\\ \nonumber && - \phi_{\mu\nu}^{\Xi^{-}} \equiv - \Delta_{\mu\nu}^9 = 2 \epsilon^{abc} d^{T}_a C \sigma_{\mu\nu} s_b \gamma_5 s_c
\\ \nonumber && ~~~~~~~~ - {i} \epsilon^{abc} d^{T}_a C \gamma_\nu s_b \gamma_\mu \gamma_5 s_c + {i} \epsilon^{abc} d^{T}_a C \gamma_\mu s_b \gamma_\nu \gamma_5 s_c
\\ \nonumber && ~~~~~~~~ + \epsilon^{abc} s^{T}_a C \sigma_{\mu\nu} s_b \gamma_5 d_c - {i\over2} \epsilon^{abc} s^{T}_a C \gamma_\nu s_b \gamma_\mu \gamma_5 d_c
\\ \nonumber && ~~~~~~~~ + {i\over2} \epsilon^{abc} s^{T}_a C \gamma_\mu s_b \gamma_\nu \gamma_5 d_c \, ,
\\ \nonumber && -{1\over\sqrt3} \phi_{\mu\nu}^{\Omega} \equiv -{1\over\sqrt3} \Delta_{\mu\nu}^{10} = \epsilon^{abc} s^{T}_a C \sigma_{\mu\nu} s_b \gamma_5 s_c \\ \nonumber && ~~~~~~~~ - {i\over2} \epsilon^{abc} s^{T}_a C \gamma_\nu s_b \gamma_\mu \gamma_5 s_c + {i\over2} \epsilon^{abc} s^{T}_a C \gamma_\mu s_b \gamma_\nu \gamma_5 s_c \, .
\end{eqnarray}

%
\section{SVZ sum rules}\label{sec:svz}
%

For the past decades QCD sum rule has proven to be a very powerful and successful non-perturbative method~\cite{Shifman:1978bx,Reinders:1984sr}. In sum rule analyses, we consider two-point correlation functions:
%
\begin{equation}
\Pi(q^2) \, \equiv \, i \int d^4x e^{iqx} \langle 0 | T J(x) { \bar J} (0) | 0 \rangle \, , \label{def:pi}
\end{equation}
%
where $J(x)$ is an interpolating field (current) coupling to a baryon state. The Lorentz structure can be simplified to be:
%
\begin{equation}
\Pi(q^2) = q\!\!\!\slash \times \Pi^{(1)}(q^2) + {\bf 1} \times \Pi^{(0)}(q^2) \, .
\label{def:pi1}
\end{equation}
%

We compute $\Pi(q^2)$ in the operator product expansion (OPE) of QCD up to certain order in the expansion, which is then matched with a hadronic parametrization to extract information about hadron properties. At the hadron level, we express the correlation function in the form of the dispersion relation with a spectral function:
%
\begin{equation}
\Pi^{(1)}(q^2)={1\over\pi}\int^\infty_{s_<}\frac{{\rm Im} \Pi^{(1)}(s)}{s-q^2-i\varepsilon}ds \, , \label{eq:disper}
\end{equation}
%
where the integration starts from the mass square of all current quarks. The imaginal part of the two-point correlation function is
%
\begin{eqnarray}
{\rm Im} \Pi^{(1)}(s) & \equiv & \pi \sum_n\delta(s-M^2_n)\langle 0|\eta|n\rangle\langle n|{\eta^\dagger}|0\rangle \, . \label{eq:rho}
\end{eqnarray}
%
For the second equation, as usual, we adopt a parametrization of one pole dominance for the ground state $Y$ $\big(\langle 0 | \eta(0) | Y \rangle \equiv f_Y v(p)$; where $v(p)$ is the Dirac spinor$\big)$ and a continuum contribution. The sum rule analysis is then performed after the Borel transformation of the two expressions of the correlation function, (\ref{def:pi}) and (\ref{eq:disper})
%
\begin{equation}
\Pi^{(all)}(M_B^2)\equiv\mathcal{B}_{M_B^2}\Pi^{(1)}(p^2) = {1\over\pi} \int^\infty_{s_<} e^{-s/M_B^2} {\rm Im} \Pi^{(1)}(s) ds \, . \label{eq:borel}
\end{equation}
%
Assuming the contribution from the continuum states can be approximated well by the spectral density of OPE above a threshold value $s_0$ (duality), we arrive at the sum rule equation
%
\begin{eqnarray}
f^2_Y e^{-M_Y^2/M_B^2} &=& \Pi(s_0, M_B^2) \equiv {1\over\pi} \int^{s_0}_{s_<} e^{-s/M_B^2} {\rm Im} \Pi^{(1)}(s) ds \label{eq:fin} \, .
\end{eqnarray}
%
Differentiating Eq.~(\ref{eq:fin}) with respect to $1 / M_B^2$ and dividing it by Eq. (\ref{eq:fin}), finally we obtain
%
\begin{equation}
M^2_Y = \frac{\frac{\partial}{\partial(-1/M_B^2)}\Pi(s_0, M_B^2)}{\Pi(s_0, M_B^2)} \, .
\end{equation}
%

In the following, we study Eq.~(\ref{eq:fin}) as functions of the two parameters, the Borel mass $M_B$ and the threshold value $s_0$. We have performed the OPE calculation up to dimension ten. For example, the explicit form of Eq.~(\ref{eq:fin}) for the baryon field $J^{(\mathbf 3,\mathbf{\bar 3}) \rightarrow \mathbf{1}_f}$($= \eta^{\Lambda_1} = \Lambda$) coupling to the flavor singlet baryon $\Lambda_1$ is
%
\begin{eqnarray}
&& f^2_{\Lambda_1} e^{-M_{\Lambda_1}^2/M_B^2} = \Pi^{\Lambda_1}(s_0, M_B^2)
\\ \nonumber &=& \int^{s_0}_{s_<} e^{-s/M_B^2} \Big ( {1 \over 32 \pi^4} s^2 - {m_s^2 \over 4 \pi^4} s + \big( {\langle g_s^2 GG \rangle \over 64 \pi^4} + {2 m_s \langle \bar q q \rangle \over 3 \pi^2}
\\ \nonumber && + { m_s \langle \bar s s \rangle \over 2 \pi^2} \big) \Big ) ds
- {8 \langle \bar q q \rangle^2 \over 9} - {16 \langle \bar q q \rangle \langle \bar s s \rangle \over 9} - {m_s \langle g_s \bar q G q \rangle \over 6 \pi^2}
\\ \nonumber && - {m_s^2 \langle g_s^2 GG \rangle \over 96 \pi^4} + {1 \over M_B^2} \Big ( {2 \langle \bar q q \rangle \langle g_s \bar q G q \rangle \over 9} + { 2 \langle \bar s s \rangle \langle g_s \bar q G q \rangle \over 9}
\\ \nonumber && + { 2 \langle \bar q q \rangle \langle g_s \bar s G s \rangle \over 9} + {m_s \langle \bar s s \rangle \langle g_s^2 GG \rangle \over 144 \pi^2} + {8 m_s^2 \langle \bar q q \rangle^2 \over 9} \Big ) \, .
\end{eqnarray}
%
We have calculated the OPEs for all the three-quark baryon fields, and the results are shown in Appendix~\ref{app:ope}.

The two-point correlation functions $\Pi^{(\mathbf 6,\mathbf{3})}_{\mu\nu}(s_0, M_B^2)$ for $[(\mathbf 6,\mathbf 3) \oplus (\mathbf 3, \mathbf 6)]$ baryons contain two free Lorentz indices $\mu$ and $\nu$, which come from their Rarita-Schwinger relevant interpolating fields, such as the octet fields $\psi_\mu^{p}$ and ${\bar \psi}_\nu^{p}$ coupling to proton:
%
\begin{equation}
\Pi^p_{\mu\nu}(q^2) \, \equiv \, i \int d^4x e^{iqx} \langle 0 | T \psi^p_\mu(x) { \bar \psi }^p_\nu (0) | 0 \rangle \, ,
\end{equation}
%
To simplify our calculations, we contract these free Lorentz indices. For example, we use the following simplified two-point correlation for the proton field $\psi_\mu^{p}$:
%
\begin{equation}
\Pi^p(q^2) \, \equiv \, \Pi^p_{\mu\mu}(q^2) \, = \, i \int d^4x e^{iqx} \langle 0 | T \psi^p_\mu(x) { \bar \psi }^p_\mu (0) | 0 \rangle \, .
\end{equation}
%
We do the same procedures for baryon fields belonging to the chiral representation $[(\mathbf {10},\mathbf 1)\oplus(\mathbf 1,\mathbf {10})]$. Here we need to contract four Lorentz indices:
%
\begin{eqnarray}\label{eq:twocontraction}
\Pi^{(\mathbf{10}, \mathbf 1)}(q^2) &\equiv& \Pi^{(\mathbf{10}, \mathbf 1)}_{\mu\nu,\mu\nu}(q^2)
\\ \nonumber &=& i \int d^4x e^{iqx} \langle 0 | T \phi(x)_{\mu\nu} { \bar \phi }_{\mu\nu} (0) | 0 \rangle \, .
\end{eqnarray}
%

\section{OPE in the $SU(3)_F$ Limit}
\label{sec:ope}

In this section we choose $m_u = m_d = m_s = 0$, $\langle \bar s s \rangle = \langle \bar q q \rangle$ and $\langle g_s \bar s G s \rangle = \langle g_s \bar q G q \rangle$, i.e., we work in the $SU(3)$ limit to see the simplified structures of OPEs. We find that the fields belonging to the same flavor multiplet lead to a same result in the QCD sum rule (except an universal factor which does not affect the final result). However, the fields belonging to the same chiral multiplet but different flavor multiplets lead to different results. Their gluon condensates are still the same, but their quark condensates are different. The latter violate the chiral symmetry, and make one chiral multiplet break to several flavor multiplets. The QCD sum rules for the baryon fields belonging to each chiral (flavor) multiplet are
\begin{eqnarray}\nonumber
&& f^2_{(\mathbf 3,\mathbf{\bar 3})\rightarrow\mathbf{1}_f} e^{-M_{(\mathbf 3,\mathbf{\bar 3})\rightarrow\mathbf{1}_f}^2/M_B^2} = \Pi^{(\mathbf 3,\mathbf{\bar 3})\rightarrow\mathbf{1}_f}(s_0, M_B^2)
\\ \nonumber &=& \int^{s_0}_{s_<} e^{-s/M_B^2} \Big ( {1 \over 32 \pi^4} s^2 + {\langle g_s^2 GG \rangle \over 64 \pi^4} \Big ) ds
\\ \nonumber && ~~~~~~~~~~~~~~~~~~~~~~~ - {8 \langle \bar q q \rangle^2 \over 3} + {2\langle \bar q q \rangle \langle g_s \bar q G q \rangle \over 3} {1 \over M_B^2} \, ,
\\ \nonumber && f^2_{(\mathbf 3,\mathbf{\bar 3})\rightarrow\mathbf{8}_f} e^{-M_{(\mathbf 3,\mathbf{\bar 3})\rightarrow\mathbf{8}_f}^2/M_B^2} = \Pi^{(\mathbf 3,\mathbf{\bar 3})\rightarrow\mathbf{8}_f}(s_0, M_B^2)
\\ \nonumber &=& \int^{s_0}_{s_<} e^{-s/M_B^2} \Big ( {1 \over 32 \pi^4} s^2 + {\langle g_s^2 GG \rangle \over 64 \pi^4} \Big ) ds
\\ \nonumber && ~~~~~~~~~~~~~~~~~~~~~~~ + {4 \langle \bar q q \rangle^2 \over 3} + {\langle \bar q q \rangle \langle g_s \bar q G q \rangle \over 3} {1 \over M_B^2} \, ,
\\ \nonumber && f^2_{(\mathbf 8,\mathbf{1})\rightarrow\mathbf{8}_f} e^{-M_{(\mathbf 8,\mathbf{1})\rightarrow\mathbf{8}_f}^2/M_B^2} = \Pi^{(\mathbf 8,\mathbf{1})\rightarrow\mathbf{8}_f}(s_0, M_B^2)
\\ \nonumber &=& \int^{s_0}_{s_<} e^{-s/M_B^2} \Big ( {3 \over 64 \pi^4} s^2 + { 3 \langle g_s^2 GG \rangle \over 128 \pi^4} \Big ) ds \, ,
\\ && f^2_{(\mathbf 6,\mathbf{3})\rightarrow\mathbf{8}_f} e^{-M_{(\mathbf 6,\mathbf{3})\rightarrow\mathbf{8}_f}^2/M_B^2} = \Pi^{(\mathbf 6,\mathbf{3})\rightarrow\mathbf{8}_f}(s_0, M_B^2)
\\ \nonumber &\equiv& \Pi^{(\mathbf 6,\mathbf{3})\rightarrow\mathbf{8}_f}_{\mu\mu}(s_0, M_B^2)
\\ \nonumber &=& \int^{s_0}_{s_<} e^{-s/M_B^2} \Big ( {9 \over 256 \pi^4} s^2 - {15\langle g_s^2 GG \rangle \over 512 \pi^4} \Big ) ds
\\ \nonumber && ~~~~~~~~~~~~~~~~~~~~~~~ - {9 \langle \bar q q \rangle^2 \over 2 } - {17 \langle \bar q q \rangle \langle g_s \bar q G q \rangle \over 8} {1 \over M_B^2} \, ,
\\ \nonumber && f^2_{(\mathbf 6,\mathbf{3})\rightarrow\mathbf{10}_f} e^{-M_{(\mathbf 6,\mathbf{3})\rightarrow\mathbf{10}_f}^2/M_B^2} = \Pi^{(\mathbf 6,\mathbf{3})\rightarrow\mathbf{10}_f}(s_0, M_B^2)
\\ \nonumber &\equiv& \Pi^{(\mathbf 6,\mathbf{3})\rightarrow\mathbf{10}_f}_{\mu\mu}(s_0, M_B^2)
\\ \nonumber &=& \int^{s_0}_{s_<} e^{-s/M_B^2} \Big ( {9 \over 256 \pi^4} s^2 - {15\langle g_s^2 GG \rangle \over 512\pi^4} \Big ) ds
\\ \nonumber && ~~~~~~~~~~~~~~~~~~~~~~~ + {9 \langle \bar q q \rangle^2 } + {21 \langle \bar q q \rangle \langle g_s \bar q G q \rangle \over 4} {1 \over M_B^2} \, ,
\\ \nonumber && f^2_{(\mathbf {10},\mathbf{1})\rightarrow\mathbf{10}_f} e^{-M_{(\mathbf {10},\mathbf{1})\rightarrow\mathbf{10}_f}^2/M_B^2} = \Pi^{(\mathbf {10},\mathbf{1})\rightarrow\mathbf{10}_f}(s_0, M_B^2)
\\ \nonumber &\equiv& \Pi^{(\mathbf {10},\mathbf{1})\rightarrow\mathbf{10}_f}_{\mu\nu,\mu\nu}(s_0, M_B^2)
\\ \nonumber &=& {9 \langle \bar q q \rangle \langle g_s \bar q G q \rangle \over 4} {1 \over M_B^2} \, .
\end{eqnarray}

From these expressions and the detailed OPE shown in Appendix~\ref{app:ope}, we find that the baryon fields belonging to the same chiral representation
have the same leading term, especially for the singlet and octet baryons belonging to the chiral representation $[(\mathbf 3,\mathbf{\bar 3}) \oplus (\mathbf {\bar 3},\mathbf{3})]$ as well as the octet and decuplet baryons belonging to $[(\mathbf 6,\mathbf 3) \oplus (\mathbf 3, \mathbf 6)]$.
This suggests that the normalizations used in Ref.~\cite{Chen:2009sf} are well-defined.

We also find that the sum rule of the baryon fields belonging to the chiral representation $(\mathbf 8,\mathbf 1)\oplus(\mathbf 1,\mathbf 8)$ contain only the gluon condensate. This indicates that the $(\mathbf 8,\mathbf 1)\oplus(\mathbf 1,\mathbf 8)$-plet is impervious to chiral symmetry breaking which fact makes it unique among baryon fields. The baryon fields belonging to $(\mathbf {10},\mathbf 1)\oplus(\mathbf 1,\mathbf {10})$ have OPEs containing only one condensate which is too simple to perform QCD sum rule analysis. This may be because that we have contracted all the four free Lorentz parameters as shown in Eq.~(\ref{eq:twocontraction}).

\section{Numerical Analysis}
\label{sec:numerical}

To perform the numerical analysis, we use the following values for the condensates and other parameters, which correspond to the energy scale of 1 GeV~\cite{Yang:1993bp,Beringer:1900zz,Narison:2002pw,Gimenez:2005nt,Jamin:2002ev,Ioffe:2002be,Ovchinnikov:1988gk,colangelo}:
%
\begin{eqnarray}
\nonumber &&\langle\bar qq \rangle=-(0.240 \pm 0.010)^3 \mbox{ GeV}^3\, ,
\\
\nonumber &&\langle\bar ss\rangle=-(0.8\pm 0.1)\times(0.240 \pm 0.010)^3 \mbox{
GeV}^3\, ,
\\
\nonumber &&\langle g_s^2GG\rangle =(0.48\pm 0.14) \mbox{ GeV}^4\, ,
\\ \nonumber && m_u = 2.90 \pm 0.20 \mbox{ MeV}\, ,m_d = 6.35 \pm 0.27 \mbox{ MeV}\, ,
\\
\label{condensates} &&m_s=125 \pm 20 \mbox{ MeV}\, ,
\\
\nonumber && \langle g_s\bar q\sigma G
q\rangle=-M_0^2\times\langle\bar qq\rangle\, ,
\\
\nonumber &&M_0^2=(0.8\pm0.2)\mbox{ GeV}^2\, .
\end{eqnarray}
%
As usual we assume the vacuum saturation for higher dimensional operators such as $\langle 0 | \bar q q \bar q q |0\rangle \sim \langle 0 | \bar q q |0\rangle \langle 0|\bar q q |0\rangle$. There is a minus sign in the definition of the mixed condensate $\langle g_s\bar q\sigma G q\rangle$, which is different from some other QCD sum rule calculations. This is because the definition of coupling constant $g_s$ is different~\cite{Yang:1993bp,Hwang:1994vp}. We note that the current quark masses of $up$ and $down$ quarks give little contributions, and so we obtain almost the same mass for baryons belonging to the same isospin multiplet. Consequently, we do not show these terms in OPEs for simplicity.

We perform the QCD sum rule analysis according to their chiral representations in the following subsections.

\subsection{Chiral Multiplet $[(\mathbf 3,\mathbf{\bar 3}) \oplus (\mathbf{\bar 3}, \mathbf 3)]$}

\begin{figure}[hbt]
\begin{center}
\scalebox{0.41}{\includegraphics{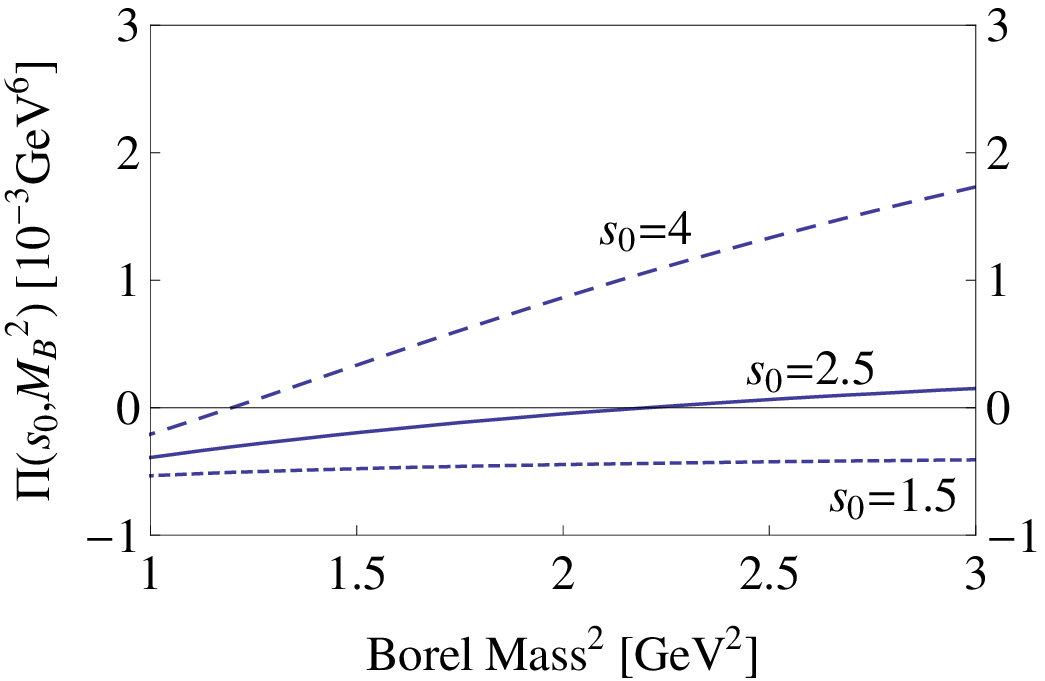}}
\scalebox{0.41}{\includegraphics{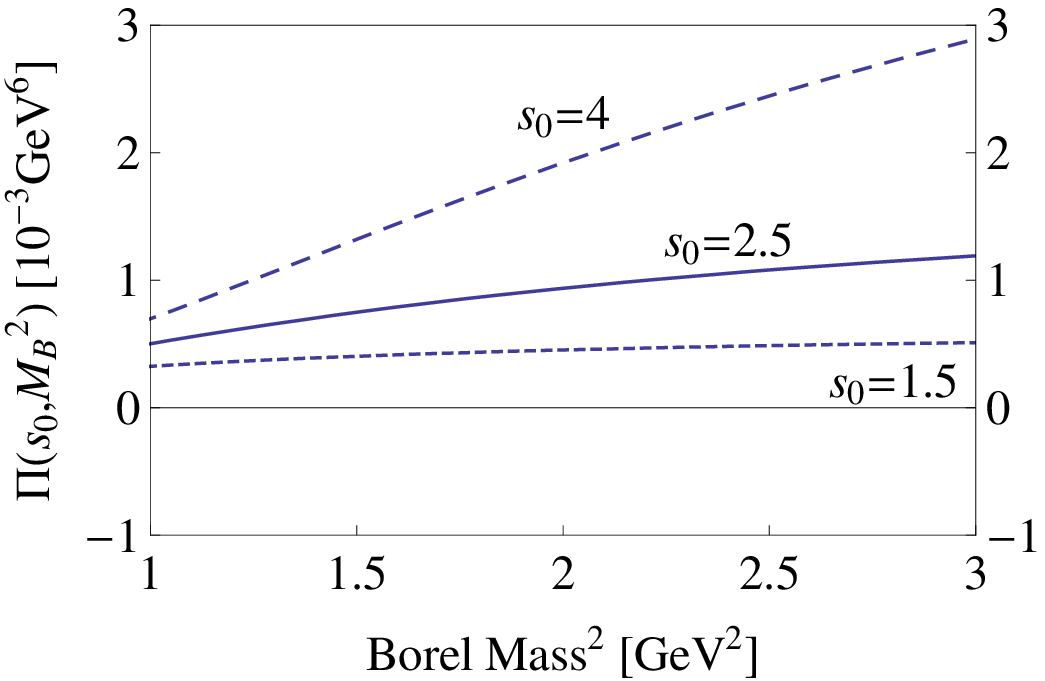}}\caption{The Borel transformed correlation functions $\Pi(s_0, M_B^2)$ calculated using the singlet baryon field $\eta^{\Lambda_1}$ (left) and the octet baryon field $\eta^{\Lambda_8}$ (right), as functions of the Borel mass $M_B$. These fields belong to the chiral representation $[(\mathbf 3,\mathbf{\bar 3}) \oplus (\mathbf{\bar 3}, \mathbf 3)]$. The short-dashed, solid and long-dashed curves are obtained by setting $s_0 = 1.5,~2.5$ and $4$ GeV$^2$, respectively.} \label{fig:PI33}
\end{center}
\end{figure}
To get a good QCD sum rule, the Borel transformed correlation function $\Pi(s_0, M_B^2)$ should be positive as defined in Eq.~(\ref{eq:fin}). However, this can not be achieved for some baryon fields. As an example, we show $\Pi(s_0, M_B^2)$ for the singlet baryon field $\eta^{\Lambda_1}$ at the left hand side of Fig.~\ref{fig:PI33}. We show it as a function of the Borel mass $M_B$, and choose $s_0 = 1.5,~2.5$ and $4$ GeV$^2$ to draw three curves. We find that only the one with $s_0 = 4$ GeV$^2$ is (almost) positive in our working region $1$ GeV$^2<M_B^2< 3$ GeV$^2$. Therefore, in order to obtain a reliable QCD sum rule, we need to use a large threshold value $s_0$, and consequently the obtained baryon mass is large. This suggests that the singlet baryon field $\eta^{\Lambda_1}$ does not support a baryon mass around 1 GeV. We would like to note that the baryons having a negative parity can also contribute in the hadronic dispersion relation (Eq.~(\ref{eq:rho})), which is, however, not considered in this paper. This was firstly discussed in Ref.~\cite{Bagan:1993ii} for heavy baryons. Then it was revisited in Refs.~\cite{Khodjamirian:2011jp,Khodjamirian:2011sp} for general baryons, emphasizing that the elimination of the background contribution due to negative-parity baryons is essential to ensure the insensitivity of baryonic sum rules to the choices of interpolating currents. This new approach can be helpful to construct the sum rules using the current $\eta^{\Lambda_1}$ as well as other baryonic currents.

\begin{figure}[hbt]
\begin{center}
\scalebox{0.34}{\includegraphics{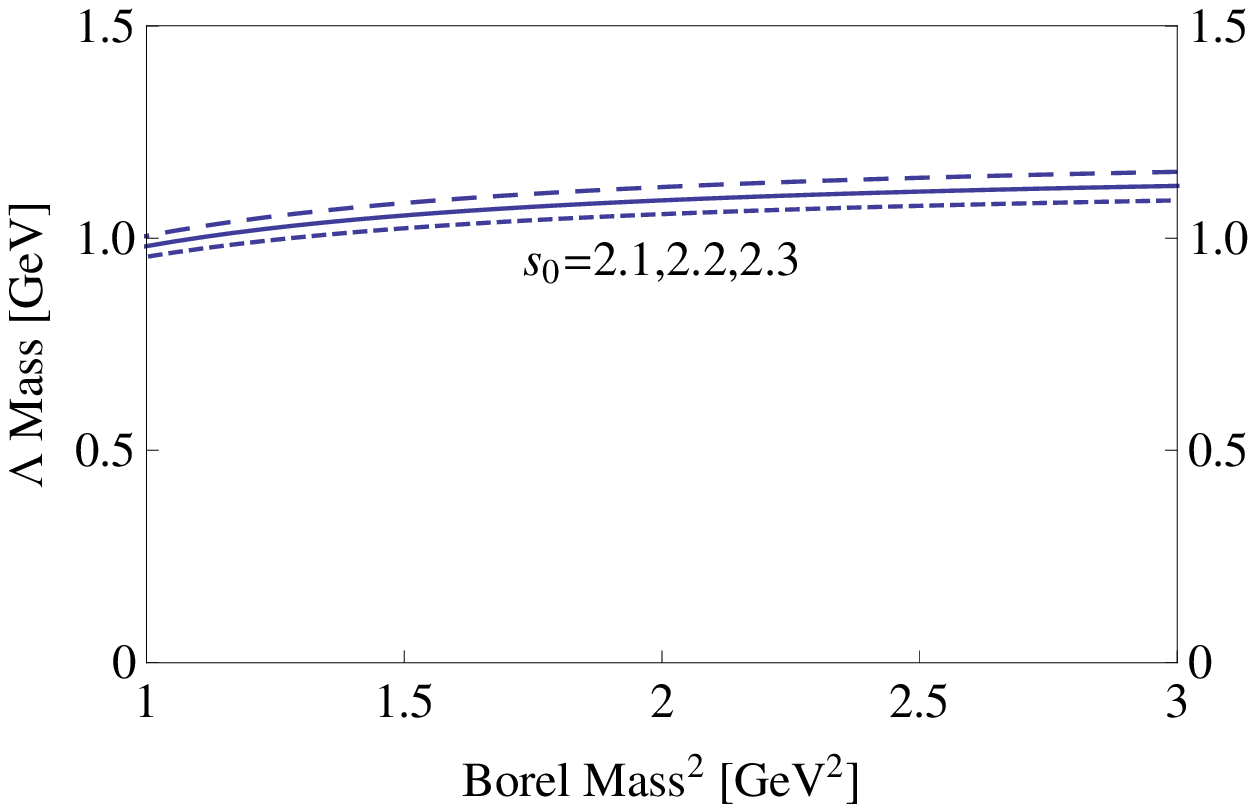}}
\scalebox{0.34}{\includegraphics{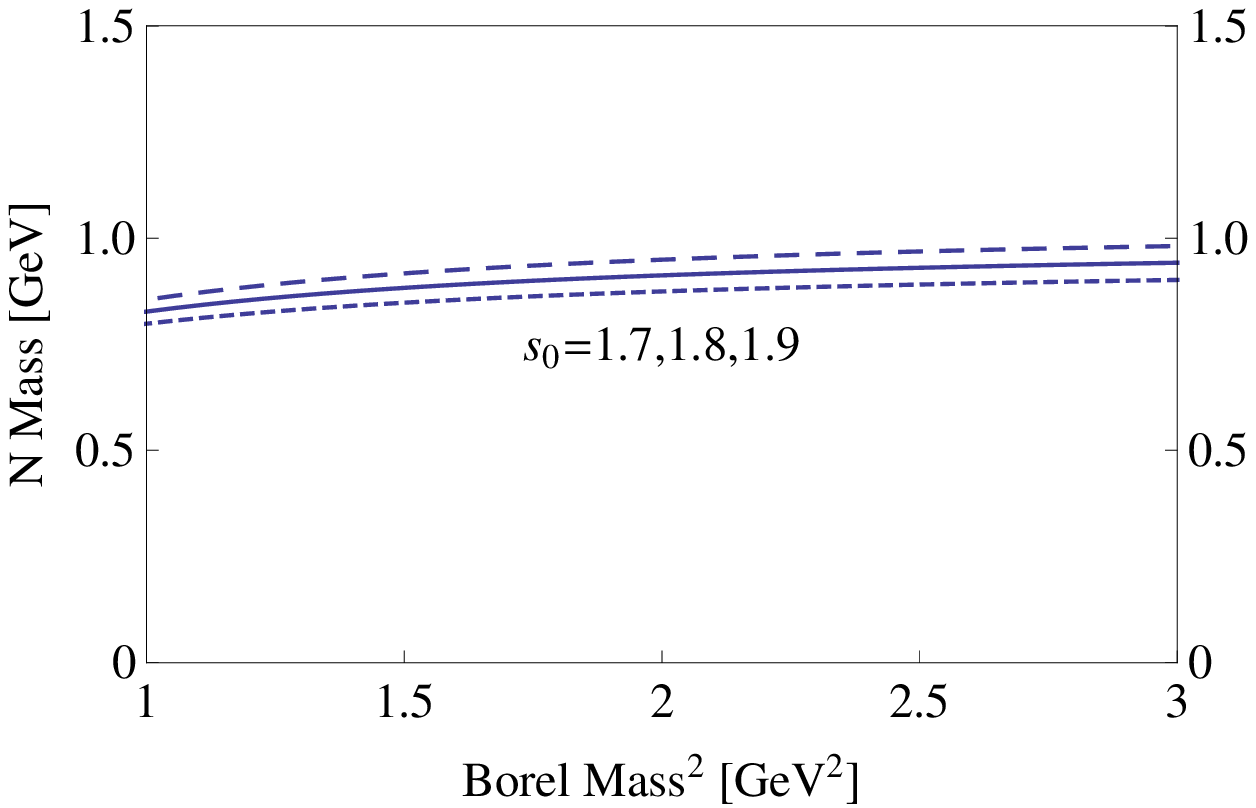}}
\scalebox{0.34}{\includegraphics{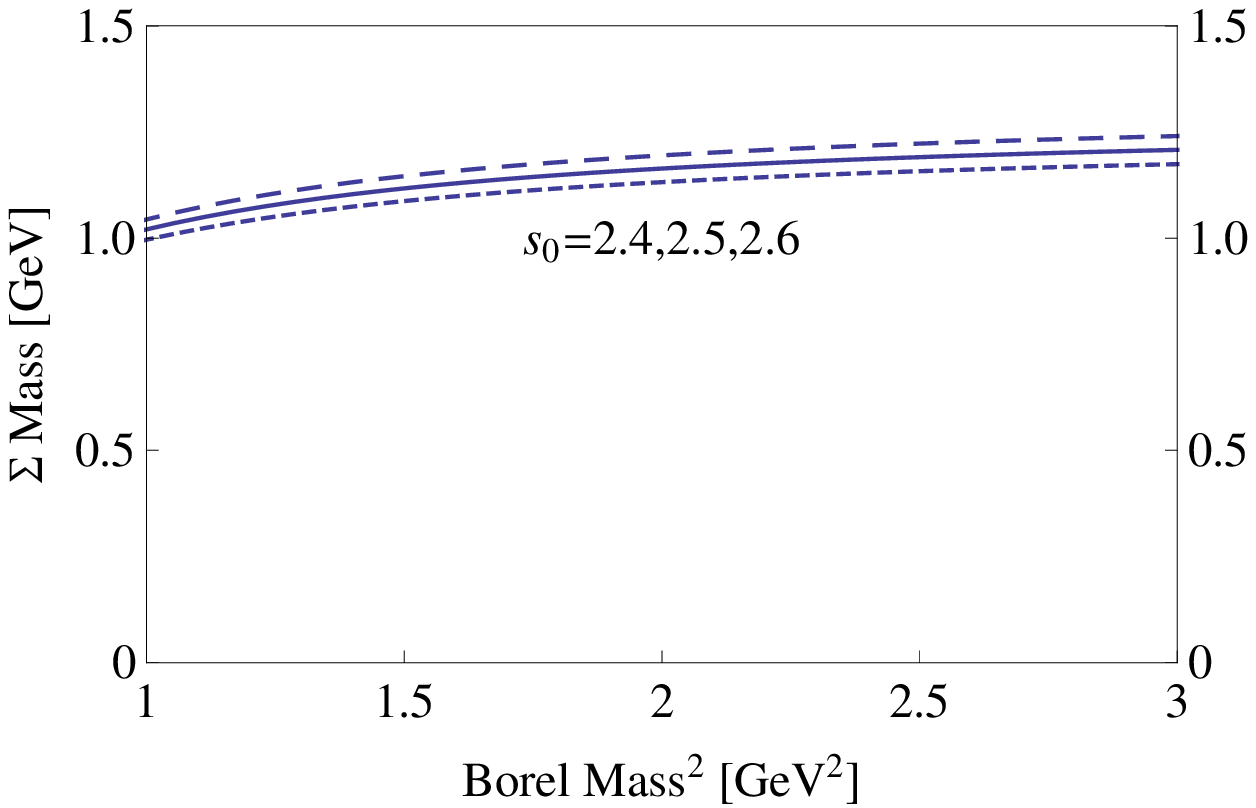}}
\scalebox{0.34}{\includegraphics{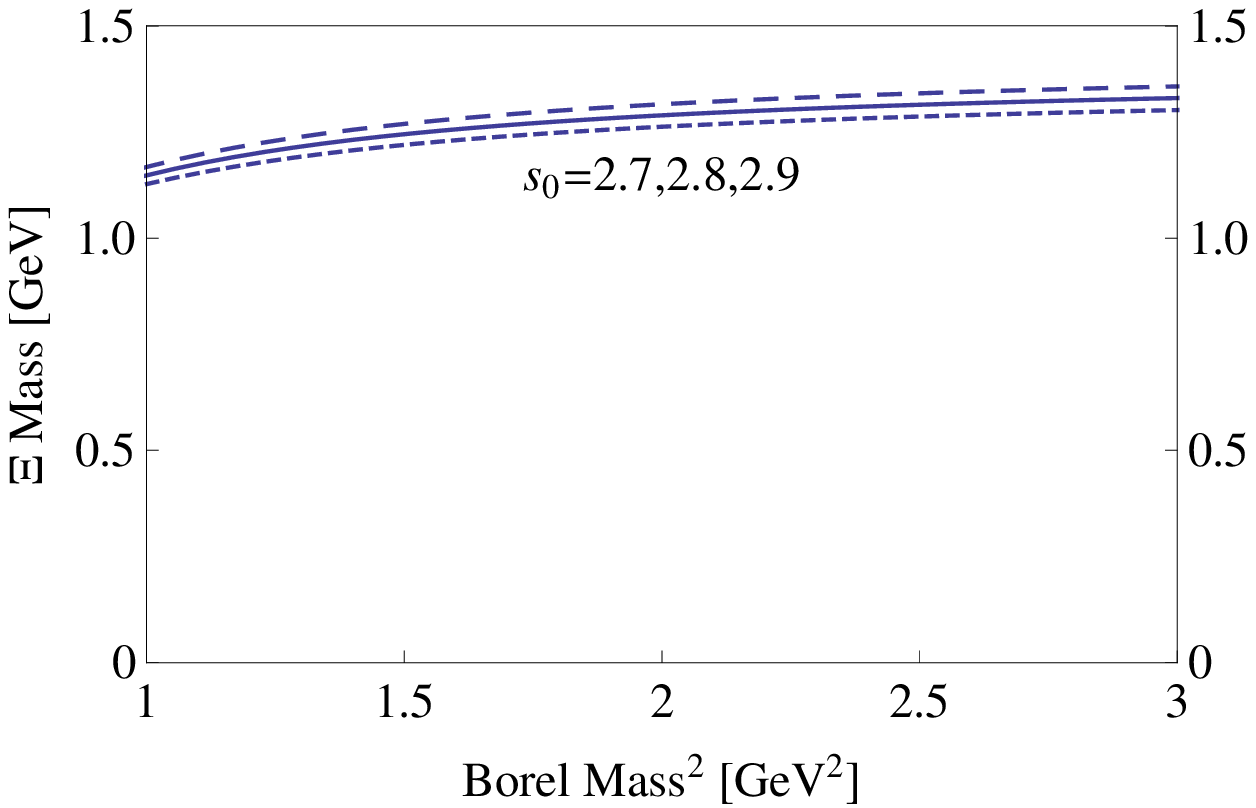}} \caption{The extracted mass of octet baryons, as functions of the Borel mass $M_B$. They belong to the chiral representation $[(\mathbf 3,\mathbf{\bar 3}) \oplus (\mathbf{\bar 3}, \mathbf 3)]$. The top-left, top-right, bottom-left and bottom-right figures are for $\Lambda_8$, $N$, $\Sigma$ and $\Xi$, respectively.} \label{fig:mass33}
\end{center}
\end{figure}
The Borel transformed correlation function $\Pi(s_0, M_B^2)$ for the octet baryon field $\eta^{\Lambda_8}$ is shown at the right hand side of Fig.~\ref{fig:PI33}. We show it as a function of the Borel mass $M_B$ and choose $s_0 = 1.5,~2.5$ and $4$ GeV$^2$. We find that all curves are positive in our working region $1.5$ GeV$^2<s_0<4$ GeV$^2$ and $1$ GeV$^2<M_B^2<3$ GeV$^2$. Using these fields, we calculate the masses of the octet baryons $\Lambda$, $N$, $\Sigma$ and $\Xi$, which are shown in Fig.~\ref{fig:mass33}. Here we do not consider the isospin breaking effects, and so the baryons belonging to the same isospin multiplet have the same mass. The obtained masses are about $m_{\Lambda}=1.10\pm0.07$~GeV, $m_N=0.93\pm0.07$~GeV, $m_{\Sigma} = 1.19\pm0.08$~GeV and $m_{\Xi}=1.31\pm0.09$~GeV. The uncertainties are obtained by changing $M_B$ in our working region $2$ GeV$^2<M_B^2< 3$ GeV$^2$, changing the $strange$ quark mass from $m_s(1$~GeV$) = 125\pm20$~MeV to $m_s(2$~GeV$) = 96.1\pm4.8$~MeV~\cite{Narison:2005ny}, and assuming that the uncertainties of $\langle \bar q q \rangle$ and $\langle g_s \bar q \sigma G q \rangle$ are about 10\%. They are compatible/consistent with the experimental data: $m_{\Lambda}=1116$~MeV, $m_p=938$~MeV, $m_{\Sigma^+} = 1189$~MeV and $m_{\Xi^0}=1315$~MeV~\cite{Nakamura:2010zzi}. We also calculate their decay constants using Eq.~(\ref{eq:fin}). The results are $f_\Lambda = 0.038 \pm 0.004$~GeV$^3$, $f_N = 0.031 \pm 0.004$~GeV$^3$, $f_\Sigma = 0.042 \pm 0.004$~GeV$^3$ and $f_\Xi = 0.048 \pm 0.005$~GeV$^3$.

\subsection{Chiral Multiplet $[(\mathbf 8,\mathbf{1}) \oplus (\mathbf{1}, \mathbf 8)]$}

\begin{figure}[hbt]
\begin{center}
\scalebox{0.6}{\includegraphics{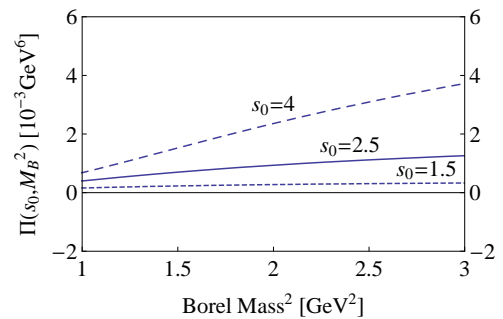}} \caption{The Borel transformed correlation function $\Pi(s_0, M_B^2)$ calculated using the octet baryon field $\xi^{\Lambda}$, as a function of the Borel mass $M_B$. This field belongs to the chiral representation $[(\mathbf 8,\mathbf{1}) \oplus (\mathbf{1}, \mathbf 8)]$. The short-dashed, solid and long-dashed curves are obtained by setting $s_0 = 1.5,~2.5$ and $4$ GeV$^2$, respectively.} \label{fig:PI81}
\end{center}
\end{figure}
\begin{figure}[hbt]
\begin{center}
\scalebox{0.34}{\includegraphics{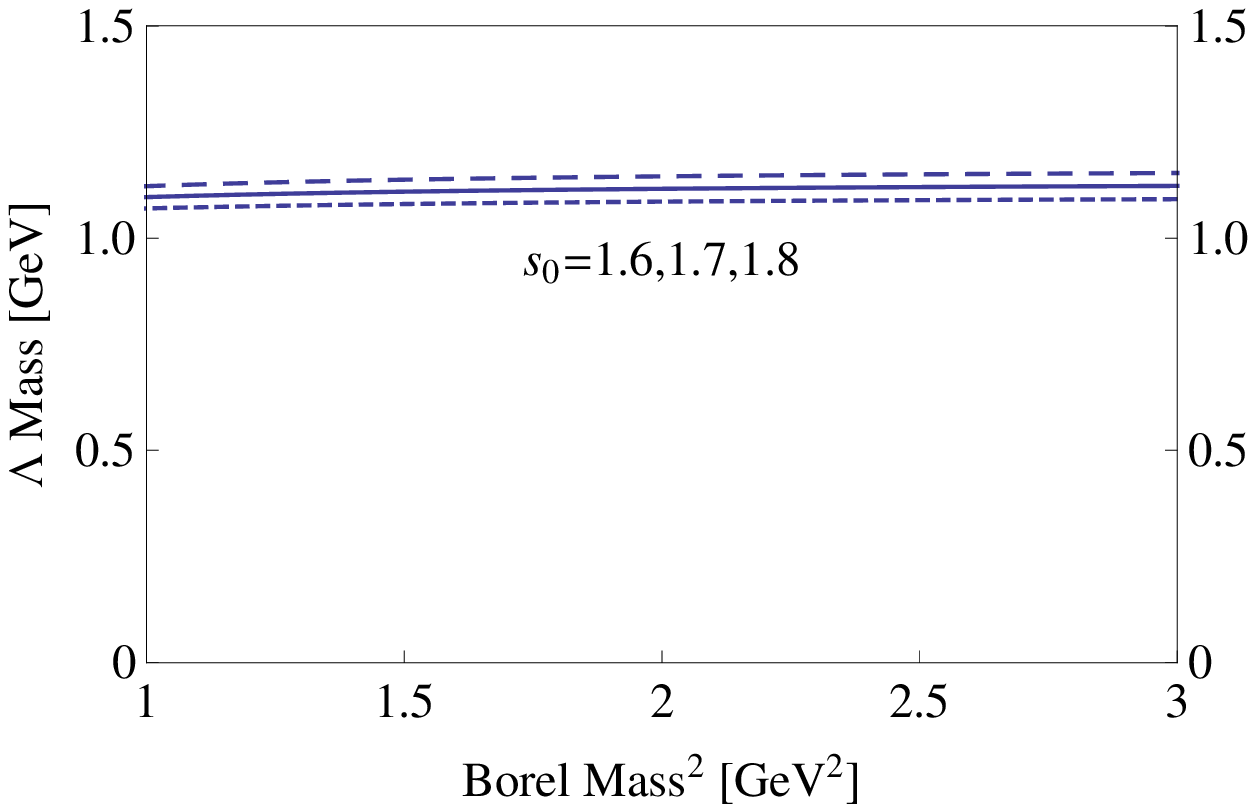}}
\scalebox{0.34}{\includegraphics{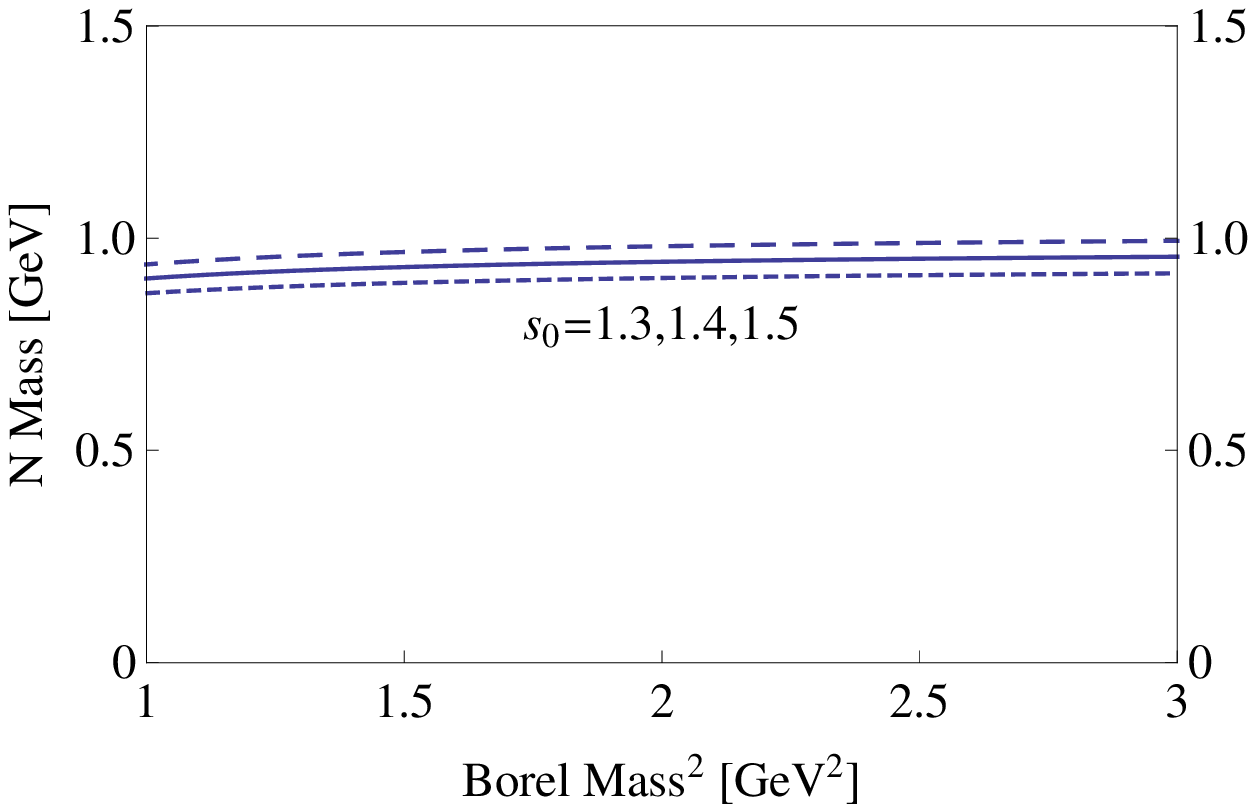}}
\scalebox{0.34}{\includegraphics{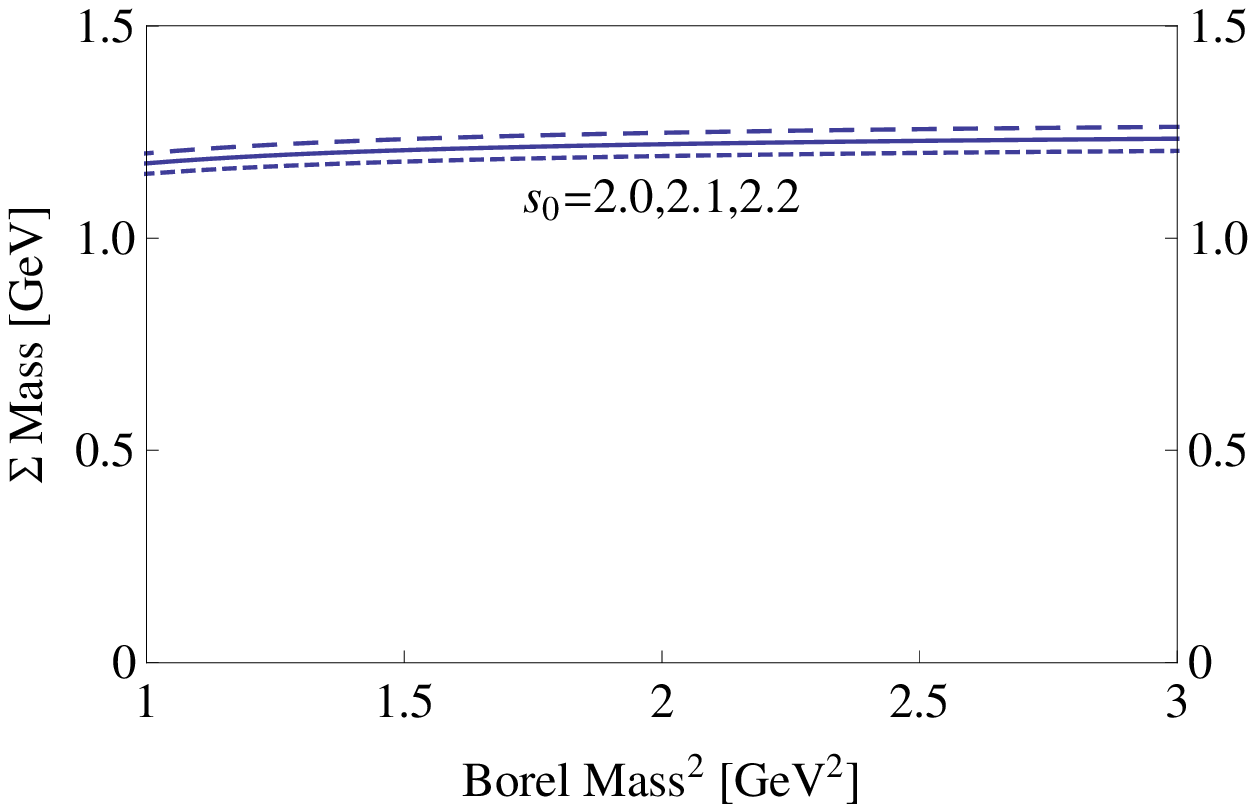}}
\scalebox{0.34}{\includegraphics{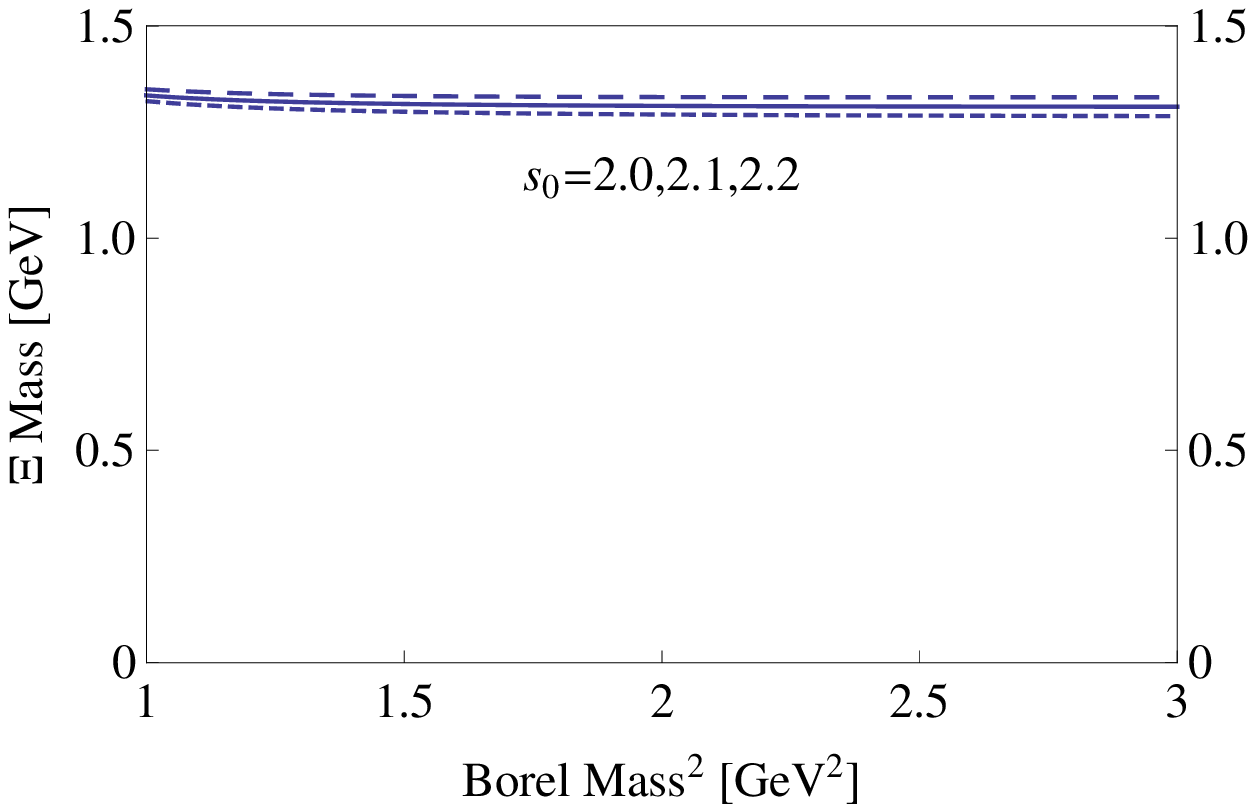}} \caption{The extracted mass of octet baryons, as functions of the Borel mass $M_B$. They belong to the chiral representation $[(\mathbf 8,\mathbf{1}) \oplus (\mathbf{1}, \mathbf 8)]$. The top-left, top-right, bottom-left and bottom-right figures are for $\Lambda_8$, $N$, $\Sigma$ and $\Xi$, respectively.} \label{fig:mass81}
\end{center}
\end{figure}
We show the Borel transformed correlation function $\Pi(s_0, M_B^2)$ for the octet baryon field $\xi^{\Lambda}$ in Fig.~\ref{fig:PI81}. We find that all curves are positive in our working region, and so we move on to perform the QCD sum rule analysis. We show the obtained masses in Fig.~\ref{fig:mass81}. The obtained masses are about $m_{\Lambda}=1.11\pm0.06$~GeV, $m_N=0.95\pm0.06$~GeV, $m_{\Sigma} = 1.23\pm0.06$~GeV and $m_{\Xi}=1.31\pm0.07$~GeV, which are compatible/consistent with the experimental data of the ground baryon masses~\cite{Nakamura:2010zzi}. The results of their decay constants are $f_\Lambda = 0.027 \pm 0.004$~GeV$^3$, $f_N = 0.024 \pm 0.004$~GeV$^3$, $f_\Sigma = 0.037 \pm 0.004$~GeV$^3$ and $f_\Xi = 0.032 \pm 0.004$~GeV$^3$.

\subsection{Chiral Multiplet $[(\mathbf 6,\mathbf 3) \oplus (\mathbf 3, \mathbf 6)]$}

\begin{figure}[hbp]
\begin{center}
\scalebox{0.41}{\includegraphics{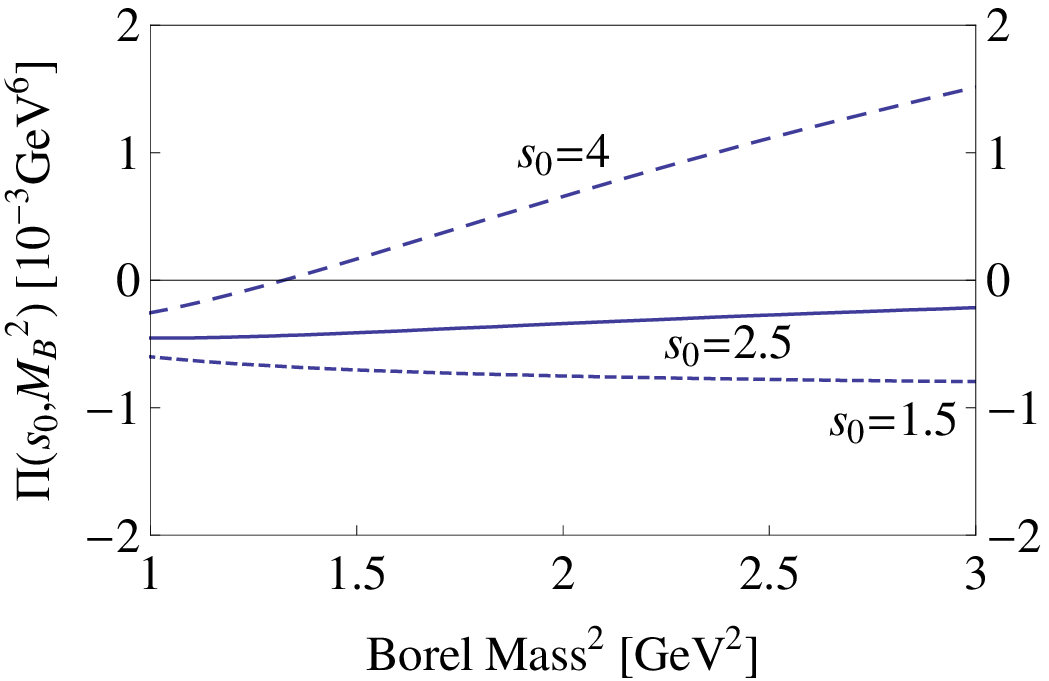}}
\scalebox{0.41}{\includegraphics{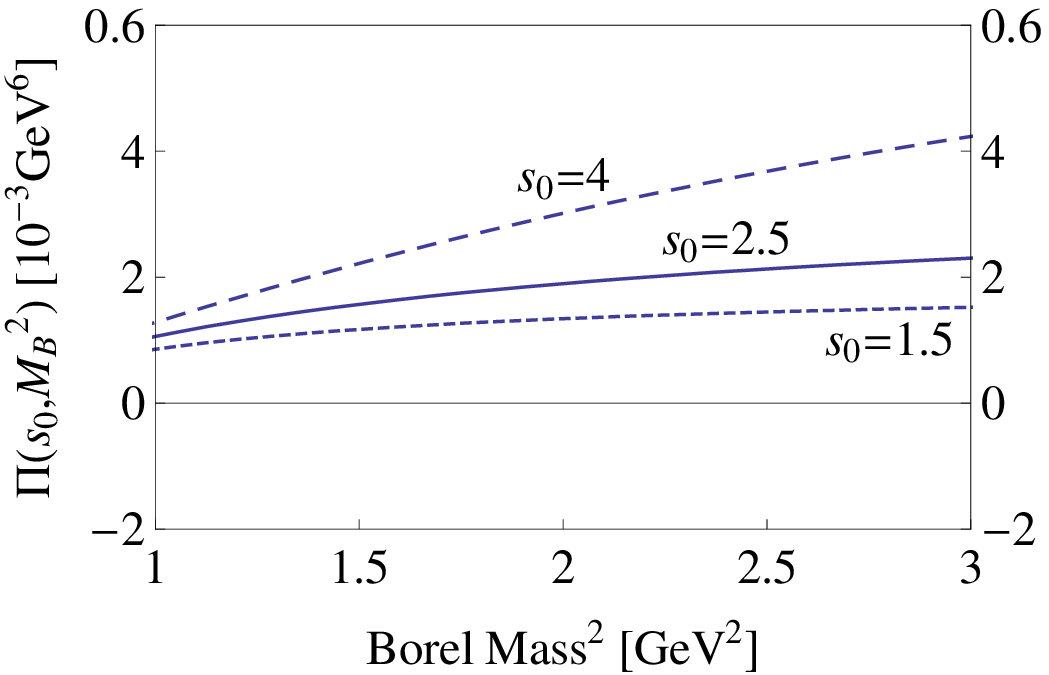}}\caption{The Borel transformed correlation functions $\Pi(s_0, M_B^2)$ calculated using the octet baryon field $\psi^{\Lambda}$ (left) and decuplet baryon field $\psi^{\Sigma^\star}$ (right), as functions of the Borel mass $M_B$. These fields belong to the chiral representation $[(\mathbf 6,\mathbf 3) \oplus (\mathbf 3, \mathbf 6)]$. The short-dashed, solid and long-dashed curves are obtained by setting $s_0 = 1.5,~2.5$ and $4$ GeV$^2$, respectively.} \label{fig:PI36}
\end{center}
\end{figure}
We show the Borel transformed correlation function $\Pi(s_0, M_B^2)$ for the octet baryon field $\psi^{\Lambda}$ at the left hand side of Fig.~\ref{fig:PI36}. We show it as a function of the Borel mass $M_B$. We find that only the one with $s_0 = 4$ GeV$^2$ is (almost) positive in our working region. Similarly to the flavor singlet field $\eta^{\Lambda_1}$, this suggests that the octet field $\psi^{\Lambda}$ does not support a baryon mass around 1 GeV. Neither do other octet members $\psi^{N}$, $\psi^{\Sigma}$ and $\psi^{\Xi}$.

\begin{figure}[hbt]
\begin{center}
\scalebox{0.34}{\includegraphics{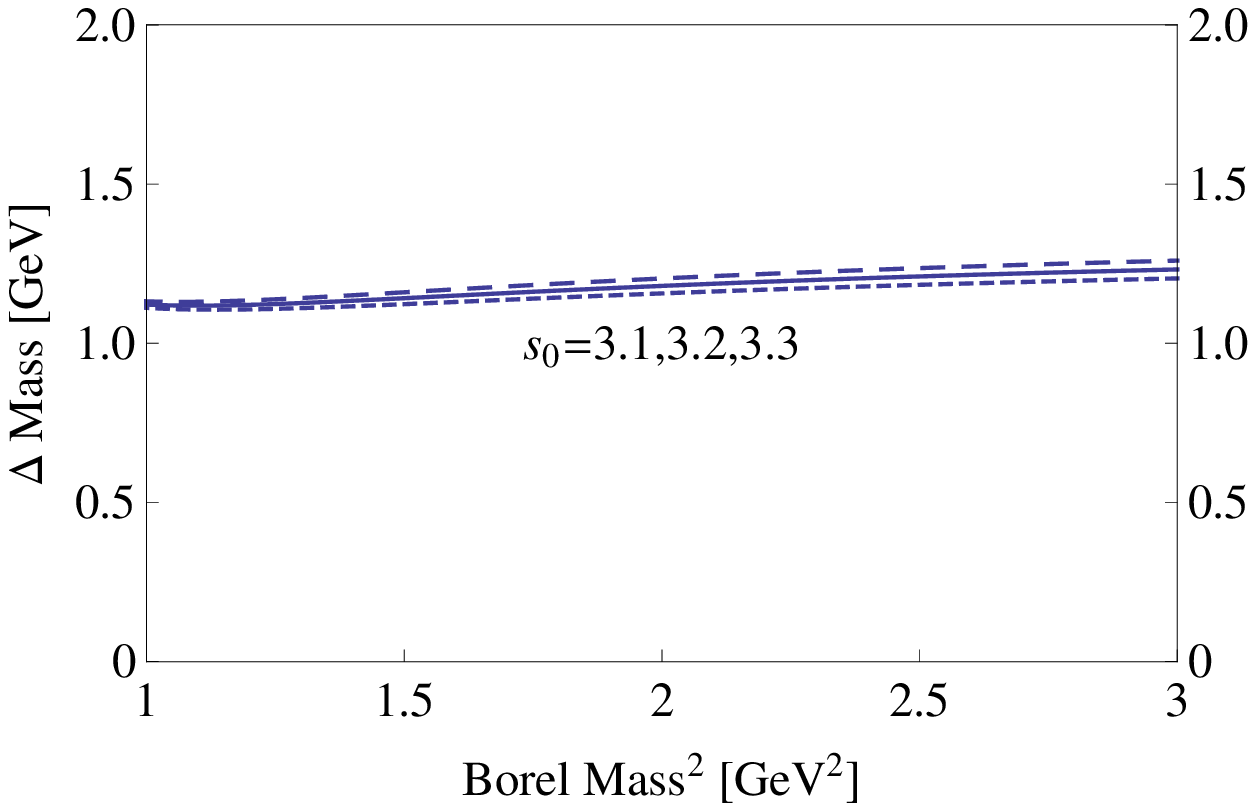}}
\scalebox{0.34}{\includegraphics{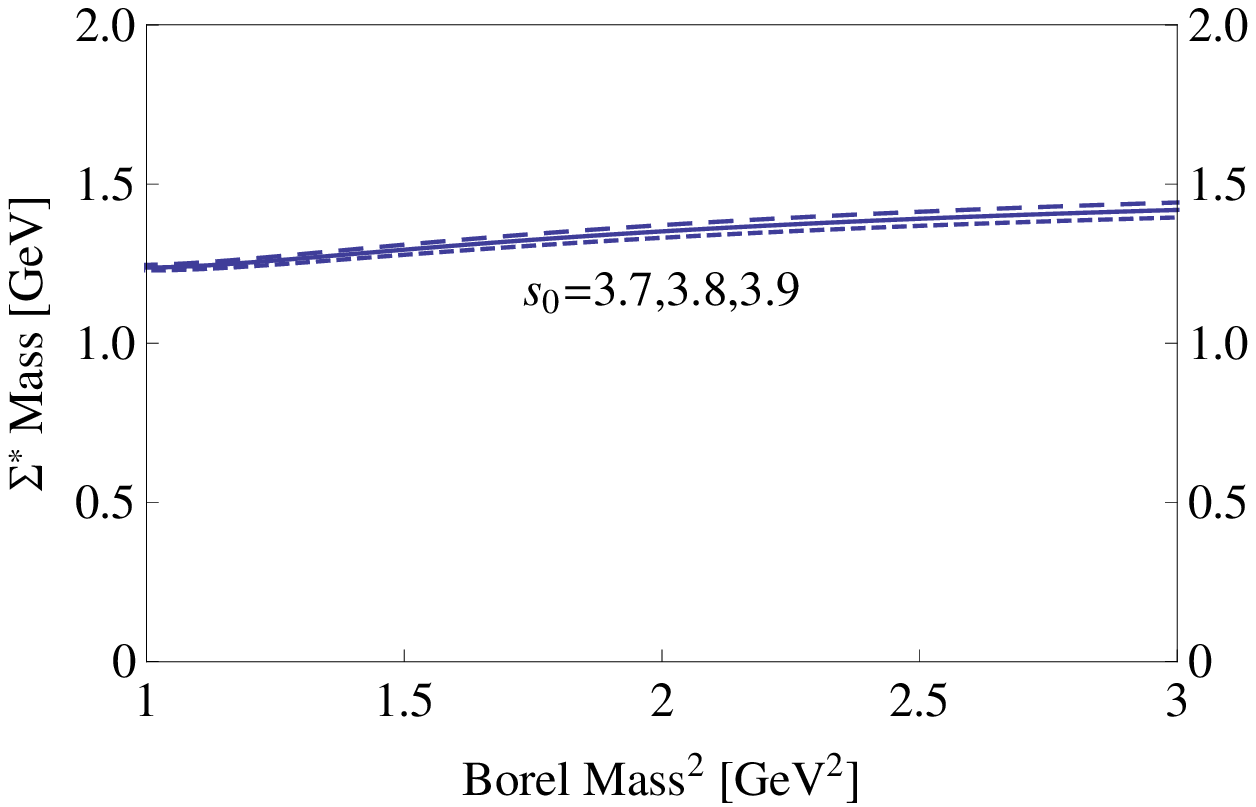}}
\scalebox{0.34}{\includegraphics{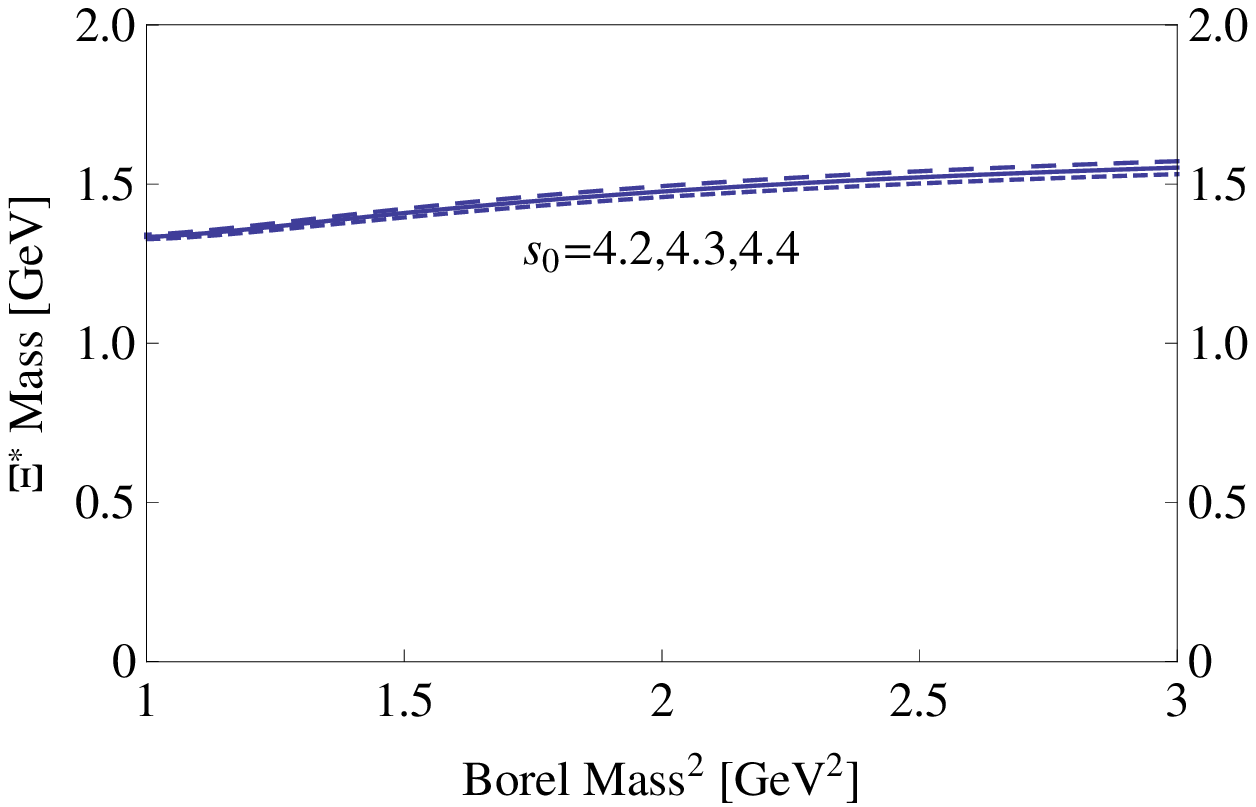}}
\scalebox{0.34}{\includegraphics{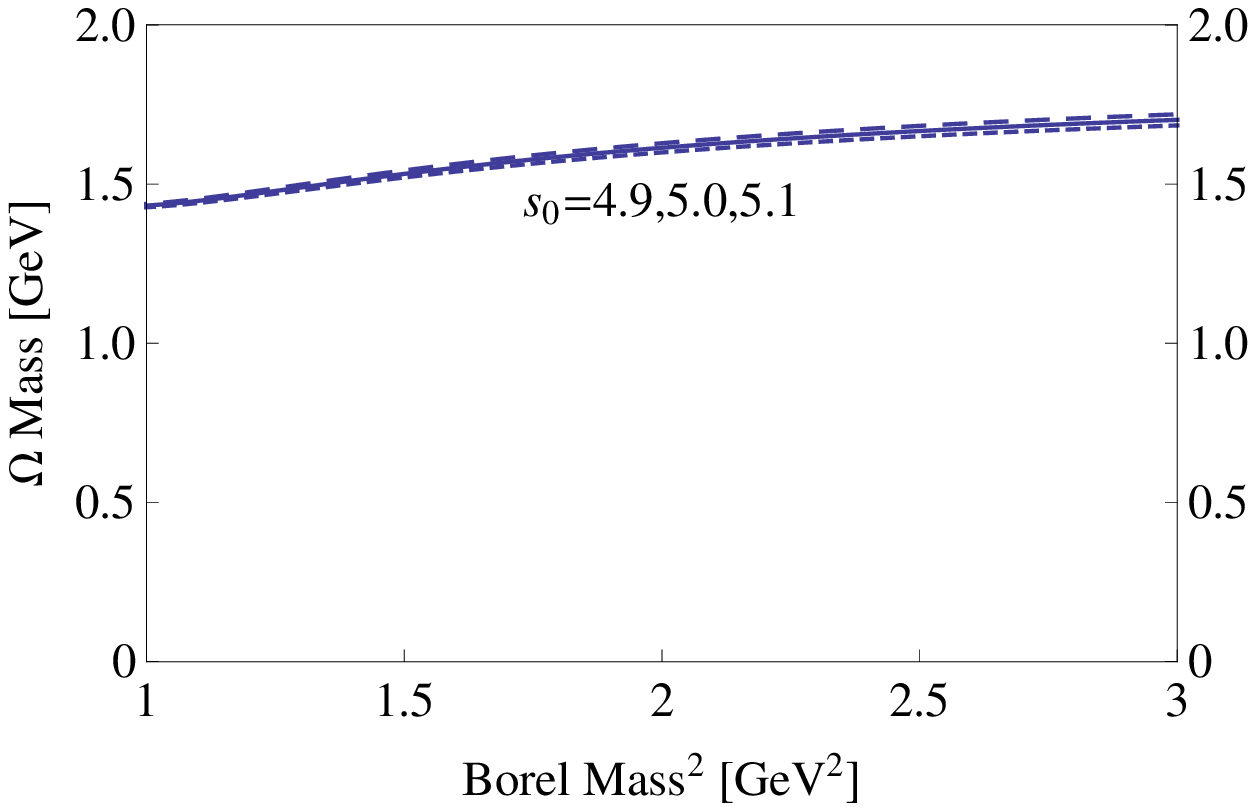}} \caption{The extracted mass of decuplet baryons, as functions of the Borel mass $M_B$. They belong to the chiral representation $[(\mathbf 6,\mathbf 3) \oplus (\mathbf 3, \mathbf 6)]$. The top-left, top-right, bottom-left and bottom-right figures are for $\Delta$, $\Sigma^\star$, $\Xi^\star$ and $\Omega$, respectively.} \label{fig:mass36}
\end{center}
\end{figure}
We show the Borel transformed correlation function $\Pi(s_0, M_B^2)$ for the decuplet baryon field $\psi^{\Sigma^\star}$ at the right hand side of Fig.~\ref{fig:PI36} as a function of the Borel mass $M_B$. We find that all curves are positive in our working region, and so we move on to calculate their masses, which are shown in Fig.~\ref{fig:mass36}. The obtained masses are about $m_{\Delta}=1.21\pm0.06$~GeV, $m_{\Sigma^\star}=1.38\pm0.07$~GeV, $m_{\Xi^\star} = 1.52\pm0.08$~GeV and $m_\Omega=1.66\pm0.09$~GeV, which are compatible/consistent with the experimental data: $m_{\Delta} = 1232$~MeV, $m_{\Sigma^\star} = 1385$~MeV, $m_{\Xi^\star} = 1530$~MeV and $m_\Omega = 1672$~MeV~\cite{Nakamura:2010zzi}. The results of their decay constants are $f_\Delta = 0.070 \pm 0.005$~GeV$^3$, $f_{\Sigma^\star} = 0.086 \pm 0.007$~GeV$^3$, $f_{\Xi^\star} = 0.10 \pm 0.01$~GeV$^3$ and $f_\Omega = 0.13 \pm 0.01$~GeV$^3$.

\subsection{Chiral Multiplet $[(\mathbf{10},\mathbf 1) \oplus (\mathbf 1, \mathbf{10})]$}

From Eqs.~(\ref{eq:ope}) and (\ref{eq:ope10}), we find that the OPEs of the decuplet baryon fields belonging to the chiral representation $[(\mathbf{10},\mathbf 1) \oplus (\mathbf 1, \mathbf{10})]$ are very simple, and so we do not perform the QCD sum rule analysis for these fields.

\section{Summary}
\label{sec:summary}

We have studied the local baryon fields according to their flavor and chiral structures in the QCD sum rule. In this work we have only considered the single baryon fields, i.e., we did not consider the mixing of baryon fields. We find that the sum rule of baryon fields belonging to the same chiral multiplet have the same leading term, which means that our normalizations in Ref.~\cite{Chen:2009sf} are well-defined. We also find that in the $SU(3)$ limit the fields belonging to the same chiral multiplet but different flavor multiplets lead to different sum rules. In the OPEs, the gluon condensates are still the same, but the quark condensates are different. The latter violate the chiral symmetry, and make one chiral multiplet break to several flavor multiplets.

We use the octet baryon fields belonging to chiral representations $[(\mathbf 3,\mathbf{\bar 3}) \oplus (\mathbf{\bar 3}, \mathbf 3)]$ and $[(\mathbf 8,\mathbf{1}) \oplus (\mathbf{1}, \mathbf 8)]$ to perform the QCD sum rule analysis, and the obtained masses are around 1.1, 0.9, 1.2 and 1.3 GeV for $\Lambda_8$, $N$, $\Sigma$ and $\Xi$, respectively, which are consistent with the experimental data of the ground octet baryon masses. To differentiate them, we would like to note that the latter one, $[(\mathbf 8,\mathbf{1}) \oplus (\mathbf{1}, \mathbf 8)]$, contains only the gluon condensate, indicating that it is impervious to chiral symmetry breaking which fact makes it unique among baryon fields. We use the decuplet baryon fields belonging to chiral representations $[(\mathbf 6,\mathbf 3) \oplus (\mathbf 3, \mathbf 6)]$, and the obtained masses are around 1.2, 1.4, 1.5 and 1.7 GeV for $\Delta$, $\Sigma^\star$, $\Xi^\star$ and $\Omega$, respectively, which are also consistent with the experimental data of the ground decuplet baryon masses. For other baryon fields, their two-point correlation functions are either non-physical (negative) or too simple to be used to perform a reliable QCD sum rule analysis. We list our conclusions in Table~\ref{tab:mass}. We have also calculated their decay constants.

We would like to note that this is our first attempt to use the method of QCD sum rule to study the baryon fields which are classified in Ref.~\cite{Chen:2008qv}. In this work only the single baryon fields are studied. However, the actual baryon is probably a mixture of several different chiral representations. In the future we shall study the mixed baryon fields, which may improve our understanding of the chiral symmetry and its spontaneously breaking.

\section*{Acknowledgments}

We would like to thank V.~Dmitra\v sinovi\' c, A.~Hosaka and S.~L.~Zhu for their careful reading and comments. This work is partly supported by the National Natural Science Foundation of China under Grant No. 11147140 and No. 11205011, and the Scientific Research Foundation for the Returned Overseas Chinese Scholars, State Education Ministry.

\onecolumn
\renewcommand{\arraystretch}{1.25}
\begin{table}[hbt]
\caption{The obtained masses using the single baryon fields. We show here whether they are consistent with the experimental data of ground baryon masses~\cite{Nakamura:2010zzi} or not.}
\begin{center}
\label{tab:mass}
\begin{tabular}{c||c|c|c}
\hline \hline
Chiral Representations & Singlet ($\Lambda_1$) & Octet ($\Lambda_8,N,\Sigma,\Xi$) & Decuplet ($\Delta,\Sigma^*,\Xi^*,\Omega$)
\\ \hline
$[(\mathbf 3,\mathbf{\bar 3}) \oplus (\mathbf{\bar 3}, \mathbf 3)]$ & $\times$ & $\surd$ & --
\\ \hline
$[(\mathbf 8,\mathbf{1}) \oplus (\mathbf{1}, \mathbf 8)]$ & -- & $\surd$ & --
\\ \hline
$[(\mathbf 6,\mathbf 3) \oplus (\mathbf 3, \mathbf 6)]$ & -- & $\times$ & $\surd$
\\ \hline
$[(\mathbf{10},\mathbf 1) \oplus (\mathbf 1, \mathbf{10})]$ & -- & -- & ?
\\ \hline \hline
\end{tabular}
\end{center}
\end{table}
\renewcommand{\arraystretch}{1}
\twocolumn

\appendix

\section{Operator Product Expansion}\label{app:ope}

In this appendix, we show the explicit forms of the operator product expansions (OPE) for the local baryon fields. The QCD sum rule for the singlet baryon field $\Lambda_1$ has been shown in the previous section and so is not shown here.

\subsection{Chiral Multiplet $[(\mathbf 3,\mathbf{\bar 3}) \oplus (\mathbf{\bar 3}, \mathbf 3)]$}

The QCD sum rules for the octet baryons belonging to the chiral representation $[(\mathbf 3,\mathbf{\bar 3}) \oplus (\mathbf{\bar 3}, \mathbf 3)]$ are
\begin{eqnarray}
\nonumber && \Pi^{\Lambda_8}(s_0, M_B^2) = f^2_{\Lambda_8} e^{-M_{\Lambda_8}^2/M_B^2}
\\ \nonumber &=& \int^{s_0}_{s_<} e^{-s/M_B^2} \Big ( {1 \over 32 \pi^4} s^2 - {m_s^2 \over 4 \pi^4} s + \big( {\langle g_s^2 GG \rangle \over 64\pi^4} - {2 m_s \langle \bar q q \rangle \over 3 \pi^2}
\\ \nonumber && + { m_s \langle \bar s s \rangle \over 2 \pi^2} \big) \Big ) ds
- {4 \langle \bar q q \rangle^2 \over 9} + {16 \langle \bar q q \rangle \langle \bar s s \rangle \over 9} - {m_s \langle g_s \bar q G q \rangle \over 6 \pi^2}
\\ \nonumber && - {m_s^2 \langle g_s^2 GG \rangle \over 48 \pi^4} + {1 \over M_B^2} \Big ( -{\langle \bar q q \rangle \langle g_s \bar q G q \rangle \over 9} + {2 \langle \bar s s \rangle \langle g_s \bar q G q \rangle \over 9}
\\ \nonumber && + {2 \langle \bar q q \rangle \langle g_s \bar s G s \rangle \over 9} + {m_s \langle \bar s s \rangle \langle g_s^2 GG \rangle \over 72 \pi^2} + {4 m_s^2 \langle \bar q q \rangle^2 \over 9} \Big ) \, ,
\\ \nonumber && \Pi^{(\mathbf 3,\mathbf{\bar 3})\rightarrow\mathbf{8}_f}(s_0, M_B^2) = f^2_{p} e^{-M_{p}^2/M_B^2} = f^2_{n} e^{-M_{n}^2/M_B^2} \, ,
\\ && \Pi^{\Sigma}(s_0, M_B^2)
\\ \nonumber &=& f^2_{\Sigma^+} e^{-M_{\Sigma^+}^2/M_B^2} = f^2_{\Sigma^0} e^{-M_{\Sigma^0}^2/M_B^2} = f^2_{\Sigma^-} e^{-M_{\Sigma^-}^2/M_B^2}
\\ \nonumber &=& \int^{s_0}_{s_<} e^{-s/M_B^2} \Big ( {s^2 \over 32 \pi^4} - {m_s^2 \over 4 \pi^4} s + \big( {\langle g_s^2 GG \rangle \over 64 \pi^4} + { m_s \langle \bar s s \rangle \over 2 \pi^2} \big) \Big ) ds
\\ \nonumber && + {4 \langle \bar q q \rangle^2 \over 3} + {1 \over M_B^2} \Big ( {\langle \bar q q \rangle \langle g_s \bar q G q \rangle \over 3} - {4 m_s^2 \langle \bar q q \rangle^2 \over 3} \Big ) \, ,
\\ \nonumber && \Pi^{\Xi}(s_0, M_B^2) = f^2_{\Xi^0} e^{-M_{\Xi^0}^2/M_B^2} = f^2_{\Xi^-} e^{-M_{\Xi^-}^2/M_B^2}
\\ \nonumber &=& \int^{s_0}_{s_<} e^{-s/M_B^2} \Big ( {1 \over 32 \pi^4} s^2 - {3 m_s^2 \over 8 \pi^4} s + {\langle g_s^2 GG \rangle \over 64 \pi^4} \Big ) ds
\\ \nonumber && + {4 \langle \bar s s \rangle^2 \over 3} - {m_s \langle g_s \bar s G s \rangle \over 4 \pi^2} - {m_s^2 \langle g_s^2 GG \rangle \over 32 \pi^4} \\ \nonumber && + {1 \over M_B^2} \Big ( {\langle \bar s s \rangle \langle g_s \bar s G s \rangle \over 3} + {m_s \langle \bar s s \rangle \langle g_s^2 GG \rangle \over 48 \pi^2} + {2 m_s^2 \langle \bar s s \rangle^2 \over 3} \Big ) \, .
\end{eqnarray}

\subsection{Chiral Multiplet $[(\mathbf 8,\mathbf{1}) \oplus (\mathbf{1}, \mathbf 8)]$}

The QCD sum rules for the octet baryons belonging to the chiral representation $[(\mathbf 8,\mathbf{1}) \oplus (\mathbf{1}, \mathbf 8)]$ are
\begin{eqnarray}
\nonumber && \Pi^{\Lambda_8}(s_0, M_B^2) = f^2_{\Lambda_8} e^{-M_{\Lambda_8}^2/M_B^2}
\\ \nonumber &=& \int^{s_0}_{s_<} e^{-s/M_B^2} \Big ( {3 \over 64 \pi^4} s^2 - {3 m_s^2 \over 8 \pi^4} s + \big( {3\langle g_s^2 GG \rangle \over 128 \pi^4}
\\ \nonumber && + {3 m_s \langle \bar s s \rangle \over 4 \pi^2} \big) \Big ) ds
 - {3 m_s^2 \langle g_s^2 GG \rangle \over 64 \pi^4} + {m_s \langle \bar s s \rangle \langle g_s^2 GG \rangle \over 32 \pi^2} {1 \over M_B^2}\, ,
\\ \nonumber && \Pi^{(\mathbf 8,\mathbf{1})\rightarrow\mathbf{8}_f}(s_0, M_B^2) = f^2_{p} e^{-M_{p}^2/M_B^2} = f^2_{n} e^{-M_{n}^2/M_B^2} \, ,
\\ && \Pi^{\Sigma}(s_0, M_B^2)
\\ \nonumber &=& f^2_{\Sigma^+} e^{-M_{\Sigma^+}^2/M_B^2} = f^2_{\Sigma^0} e^{-M_{\Sigma^0}^2/M_B^2} = f^2_{\Sigma^-} e^{-M_{\Sigma^-}^2/M_B^2}
\\ \nonumber &=& \int^{s_0}_{s_<} e^{-s/M_B^2} \Big ( {3 \over 64 \pi^4} s^2 - {3m_s^2 \over 8 \pi^4} s + \big( {3\langle g_s^2 GG \rangle \over 128 \pi^4}
\\ \nonumber && + { 3 m_s \langle \bar s s \rangle \over 4 \pi^2} \big) \Big ) ds + {m_s^2 \langle g_s^2 GG \rangle \over 64 \pi^4} - {m_s \langle \bar s s \rangle \langle g_s^2 GG \rangle \over 96 \pi^2} {1 \over M_B^2}\, ,
\\ \nonumber && \Pi^{\Xi}(s_0, M_B^2) = f^2_{\Xi^0} e^{-M_{\Xi^0}^2/M_B^2} = f^2_{\Xi^-} e^{-M_{\Xi^-}^2/M_B^2}
\\ \nonumber &=& \int^{s_0}_{s_<} e^{-s/M_B^2} \Big ( {3 \over 64 \pi^4} s^2 - {3m_s^2 \over 4 \pi^4} s + \big( {3\langle g_s^2 GG \rangle \over 128 \pi^4}
\\ \nonumber && + { 3 m_s \langle \bar s s \rangle \over 2 \pi^2} \big) \Big ) ds - {m_s^2 \langle g_s^2 GG \rangle \over 16 \pi^4} + {1 \over M_B^2} \Big ( {m_s \langle \bar s s \rangle \langle g_s^2 GG \rangle \over 24 \pi^2}
\\ \nonumber && + m_s^2 \langle \bar s s \rangle^2 \Big ) \, .
\end{eqnarray}

\subsection{Chiral Multiplet $[(\mathbf 6,\mathbf 3) \oplus (\mathbf 3, \mathbf 6)]$}

The QCD sum rules for the octet baryons belonging to the chiral representation $[(\mathbf 6,\mathbf 3) \oplus (\mathbf 3, \mathbf 6)]$ are
\begin{eqnarray}
\nonumber && \Pi^{\Lambda_8}(s_0, M_B^2) \equiv \Pi_{\mu\mu}^{\Lambda_8}(s_0, M_B^2) = f^2_{\Lambda_8} e^{-M_{\Lambda_8}^2/M_B^2}
\\ \nonumber &=& \int^{s_0}_{s_<} e^{-s/M_B^2} \Big ( {9 \over 256 \pi^4} s^2 - {9 m_s^2 \over 32 \pi^4} s + \big( - {15 \langle g_s^2 GG \rangle \over 512 \pi^4}
\\ \nonumber && + {9 m_s \langle \bar s s \rangle \over 16 \pi^2} \big) \Big ) ds - {9 \langle \bar q q \rangle^2 \over 2} - {m_s \langle g_s \bar q G q \rangle \over 4 \pi^2} + {3 m_s^2 \langle g_s^2 GG \rangle \over 128 \pi^4}
\\ \nonumber && + {1 \over M_B^2} \Big ( -{169 \langle \bar q q \rangle \langle g_s \bar q G q \rangle \over 72} + {\langle \bar s s \rangle \langle g_s \bar q G q \rangle \over 3}
\\ \nonumber && - {\langle \bar q q \rangle \langle g_s \bar s G s \rangle \over 9} - {m_s \langle \bar s s \rangle \langle g_s^2 GG \rangle \over 64 \pi^2} + {9 m_s^2 \langle \bar q q \rangle^2 \over 2} \Big ) \, ,
\\ \nonumber && \Pi^{(\mathbf 3,\mathbf 6)\rightarrow\mathbf{8}_f}(s_0, M_B^2) = f^2_{p} e^{-M_{p}^2/M_B^2} = f^2_{n} e^{-M_{n}^2/M_B^2} \, ,
\\ && \Pi^{\Sigma}(s_0, M_B^2) \equiv \Pi_{\mu\mu}^{\Sigma}(s_0, M_B^2)
\\ \nonumber &=& f^2_{\Sigma^+} e^{-M_{\Sigma^+}^2/M_B^2} = f^2_{\Sigma^0} e^{-M_{\Sigma^0}^2/M_B^2} = f^2_{\Sigma^-} e^{-M_{\Sigma^-}^2/M_B^2}
\\ \nonumber &=& \int^{s_0}_{s_<} e^{-s/M_B^2} \Big ( {9 \over 256 \pi^4} s^2 - {9 m_s^2 \over 32 \pi^4} s + \big( - {15 \langle g_s^2 GG \rangle \over 512 \pi^4}
\\ \nonumber && + { 9 m_s \langle \bar q q \rangle \over 4 \pi^2} + { 9 m_s \langle \bar s s \rangle \over 16 \pi^2} \big) \Big ) ds + {3 \langle \bar q q \rangle^2 \over 2} - {6 \langle \bar q q \rangle \langle \bar s s \rangle}
\\ \nonumber && + {21 m_s \langle g_s \bar q G q \rangle \over 16 \pi^2} + {m_s^2 \langle g_s^2 GG \rangle \over 64 \pi^4} + {1 \over M_B^2} \Big ( {25 \langle \bar q q \rangle \langle g_s \bar q G q \rangle \over 24}
\\ \nonumber && - {7 \langle \bar s s \rangle \langle g_s \bar q G q \rangle \over 4} - {17 \langle \bar q q \rangle \langle g_s \bar s G s \rangle \over 12} - {m_s \langle \bar s s \rangle \langle g_s^2 GG \rangle \over 96 \pi^2}
\\ \nonumber && - {3 m_s^2 \langle \bar q q \rangle^2 \over 2} \Big ) \, ,
\\ \nonumber && \Pi^{\Xi}(s_0, M_B^2) \equiv \Pi_{\mu\mu}^{\Xi}(s_0, M_B^2)
\\ \nonumber &=& f^2_{\Xi^0} e^{-M_{\Xi^0}^2/M_B^2} = f^2_{\Xi^-} e^{-M_{\Xi^-}^2/M_B^2}
\\ \nonumber &=& \int^{s_0}_{s_<} e^{-s/M_B^2} \Big ( {9 \over 256 \pi^4} s^2 - {27 m_s^2 \over 64 \pi^4} s + \big( - {15 \langle g_s^2 GG \rangle \over 512 \pi^4}
\\ \nonumber && + { 9 m_s \langle \bar q q \rangle \over 4 \pi^2} \big) \Big ) ds - {6 \langle \bar q q \rangle \langle \bar s s \rangle} + {3 \langle \bar s s \rangle^2 \over 2} + {15 m_s \langle g_s \bar q G q \rangle \over 16 \pi^2}
\\ \nonumber && - {21 m_s \langle g_s \bar s G s \rangle \over 32 \pi^2} + {11 m_s^2 \langle g_s^2 GG \rangle \over 256 \pi^4}
\\ \nonumber && + {1 \over M_B^2} \Big ( - {5 \langle \bar s s \rangle \langle g_s \bar q G q \rangle \over 4} - {7 \langle \bar q q \rangle \langle g_s \bar s G s \rangle \over 4}
\\ \nonumber && + {7 \langle \bar s s \rangle \langle g_s \bar s G s \rangle \over 8} - {11 m_s \langle \bar s s \rangle \langle g_s^2 GG \rangle \over 384 \pi^2}
\\ \nonumber && + {9 m_s^2 \langle \bar q q \rangle \langle \bar s s \rangle} + {3 m_s^2 \langle \bar s s \rangle^2 \over 4} \Big ) \, .
\end{eqnarray}
The QCD sum rules for the decuplet baryons belonging to the chiral representation $[(\mathbf 6,\mathbf 3) \oplus (\mathbf 3, \mathbf 6)]$ are
\begin{eqnarray}
\nonumber && \Pi^{(\mathbf 3,\mathbf 6)\rightarrow\mathbf{10}_f}(s_0, M_B^2) = f^2_{\Delta^{++}} e^{-M_{\Delta^{++}}^2/M_B^2}
\\ \nonumber &=& f^2_{\Delta^{+}} e^{-M_{\Delta^{+}}^2/M_B^2} = f^2_{\Delta^{0}} e^{-M_{\Delta^{0}}^2/M_B^2} = f^2_{\Delta^{-}} e^{-M_{\Delta^{-}}^2/M_B^2} \, ,
\\ \nonumber && \Pi^{\Sigma^*}(s_0, M_B^2) = \Pi_{\mu\mu}^{\Sigma^*}(s_0, M_B^2) = f^2_{\Sigma^{*+}} e^{-M_{\Sigma^{*+}}^2/M_B^2}
\\ \nonumber &=& f^2_{\Sigma^{*0}} e^{-M_{\Sigma^{*0}}^2/M_B^2} = f^2_{\Sigma^{*-}} e^{-M_{\Sigma^{*-}}^2/M_B^2}
\\ \nonumber &=& \int^{s_0}_{s_<} e^{-s/M_B^2} \Big ( {9 \over 256 \pi^4} s^2 - {9 m_s^2 \over 32 \pi^4} s + \big( - {15 \langle g_s^2 GG \rangle \over 512 \pi^4}
\\ \nonumber && - { 9 m_s \langle \bar q q \rangle \over 4 \pi^2} + { 9 m_s \langle \bar s s \rangle \over 16 \pi^2} \big) \Big ) ds
+ {3 \langle \bar q q \rangle^2} + {6 \langle \bar q q \rangle \langle \bar s s \rangle}
\\ \nonumber && - {21 m_s \langle g_s \bar q G q \rangle \over 16 \pi^2} + {5 m_s^2 \langle g_s^2 GG \rangle \over 256 \pi^4} + {1 \over M_B^2} \Big ( {7 \langle \bar q q \rangle \langle g_s \bar q G q \rangle \over 4}
\\ \nonumber && + {7 \langle \bar s s \rangle \langle g_s \bar q G q \rangle \over 4} + {7 \langle \bar q q \rangle \langle g_s \bar s G s \rangle \over 4} - {5 m_s \langle \bar s s \rangle \langle g_s^2 GG \rangle \over 384 \pi^2}
\\ \nonumber && - {3 m_s^2 \langle \bar q q \rangle^2} \Big ) \, ,
\\ && \Pi^{\Xi^*}(s_0, M_B^2) = \Pi_{\mu\mu}^{\Xi^*}(s_0, M_B^2)
\\ \nonumber &=& f^2_{\Xi^{*0}} e^{-M_{\Xi^{*0}}^2/M_B^2} = f^2_{\Xi^{*-}} e^{-M_{\Xi^{*-}}^2/M_B^2}
\\ \nonumber &=& \int^{s_0}_{s_<} e^{-s/M_B^2} \Big ( {9 \over 256 \pi^4} s^2 - {9 m_s^2 \over 32 \pi^4} s + \big( - {15 \langle g_s^2 GG \rangle \over 512 \pi^4}
\\ \nonumber && - { 9 m_s \langle \bar q q \rangle \over 4 \pi^2} - { 9 m_s \langle \bar s s \rangle \over 8 \pi^2} \big) \Big ) ds + {6 \langle \bar q q \rangle \langle \bar s s \rangle} + {3 \langle \bar s s \rangle^2}
\\ \nonumber && - {21 m_s \langle g_s \bar q G q \rangle \over 16 \pi^2} - {21 m_s \langle g_s \bar s G s \rangle \over 16 \pi^2} + {5 m_s^2 \langle g_s^2 GG \rangle \over 128 \pi^4}
\\ \nonumber && + {1 \over M_B^2} \Big ( {7 \langle \bar s s \rangle \langle g_s \bar q G q \rangle \over 4} + {7 \langle \bar q q \rangle \langle g_s \bar s G s \rangle \over 4} + {7 \langle \bar s s \rangle \langle g_s \bar s G s \rangle \over 4}
\\ \nonumber && - {5 m_s \langle \bar s s \rangle \langle g_s^2 GG \rangle \over 192 \pi^2} - {9 m_s^2 \langle \bar q q \rangle \langle \bar s s \rangle} + {3 m_s^2 \langle \bar s s \rangle^2 \over 4} \Big ) \, ,
\\ \nonumber && \Pi^{\Omega}(s_0, M_B^2) = \Pi_{\mu\mu}^{\Omega}(s_0, M_B^2) = f^2_{\Omega} e^{-M_{\Omega}^2/M_B^2}
\\ \nonumber &=& \int^{s_0}_{s_<} e^{-s/M_B^2} \Big ( {9 \over 256 \pi^4} s^2 + \big( - {15 \langle g_s^2 GG \rangle \over 512 \pi^4}
\\ \nonumber && - { 81 m_s \langle \bar s s \rangle \over 16 \pi^2} \big) \Big ) ds + {9 \langle \bar s s \rangle^2} - {63 m_s \langle g_s \bar s G s \rangle \over 16 \pi^2}
\\ \nonumber && + {15 m_s^2 \langle g_s^2 GG \rangle \over 256 \pi^4} + {1 \over M_B^2} \Big ( {21 \langle \bar s s \rangle \langle g_s \bar s G s \rangle \over 4}
\\ \nonumber && - {5 m_s \langle \bar s s \rangle \langle g_s^2 GG \rangle \over 128 \pi^2} - {63 m_s^2 \langle \bar s s \rangle^2 \over 4} \Big ) \, .
\end{eqnarray}

\subsection{Chiral Multiplet $[(\mathbf{10},\mathbf 1) \oplus (\mathbf 1, \mathbf{10})]$}

The QCD sum rules for the decuplet baryons belonging to the chiral representation $[(\mathbf{10},\mathbf 1) \oplus (\mathbf 1, \mathbf{10})]$ are
\begin{eqnarray}\label{eq:ope10}
\nonumber && \Pi^{(\mathbf{10},\mathbf 1)\rightarrow\mathbf{10}_f}(s_0, M_B^2) = f^2_{\Delta^{++}} e^{-M_{\Delta^{++}}^2/M_B^2}
\\ \nonumber &=& f^2_{\Delta^{+}} e^{-M_{\Delta^{+}}^2/M_B^2} = f^2_{\Delta^{0}} e^{-M_{\Delta^{0}}^2/M_B^2} = f^2_{\Delta^{-}} e^{-M_{\Delta^{-}}^2/M_B^2} \, ,
\\ \nonumber && \Pi^{\Sigma^*}(s_0, M_B^2) = \Pi_{\mu\nu,\mu\nu}^{\Sigma^*}(s_0, M_B^2) = f^2_{\Sigma^{*+}} e^{-M_{\Sigma^{*+}}^2/M_B^2}
\\ \nonumber &=& f^2_{\Sigma^{*0}} e^{-M_{\Sigma^{*0}}^2/M_B^2} = f^2_{\Sigma^{*-}} e^{-M_{\Sigma^{*-}}^2/M_B^2}
\\ \nonumber &=& - {5 m_s \langle g_s \bar q G q \rangle \over 8 \pi^2} + {1 \over M_B^2} \Big ( {7 \langle \bar q q \rangle \langle g_s \bar q G q \rangle \over 12} + {5 \langle \bar s s \rangle \langle g_s \bar q G q \rangle \over 6}
\\ \nonumber && + {5 \langle \bar q q \rangle \langle g_s \bar s G s \rangle \over 6} \Big ) \, ,
\\ && \Pi^{\Xi^*}(s_0, M_B^2) = \Pi_{\mu\nu,\mu\nu}^{\Xi^*}(s_0, M_B^2)
\\ \nonumber &=& f^2_{\Xi^{*0}} e^{-M_{\Xi^{*0}}^2/M_B^2} = f^2_{\Xi^{*-}} e^{-M_{\Xi^{*-}}^2/M_B^2}
\\ \nonumber &=& - {5 m_s \langle g_s \bar q G q \rangle \over 8 \pi^2} - {7 m_s \langle g_s \bar s G s \rangle \over 16 \pi^2} + {1 \over M_B^2} \Big ( {5 \langle \bar s s \rangle \langle g_s \bar q G q \rangle \over 6}
\\ \nonumber && + {5 \langle \bar q q \rangle \langle g_s \bar s G s \rangle \over 6} + {7 \langle \bar s s \rangle \langle g_s \bar s G s \rangle \over 12} \Big ) \, ,
\\ \nonumber && \Pi^{\Omega}(s_0, M_B^2) = \Pi_{\mu\nu,\mu\nu}^{\Omega}(s_0, M_B^2) = f^2_{\Omega} e^{-M_{\Omega}^2/M_B^2}
\\ \nonumber &=& - {27 m_s \langle g_s \bar s G s \rangle \over 16 \pi^2} + {9 \langle \bar s s \rangle \langle g_s \bar s G s \rangle \over 4} {1 \over M_B^2} \, .
\end{eqnarray}
%
%

\end{document}